\documentclass[prb,preprint,amsmath,amssymb,citeautoscript,showkeys,floatfix]{revtex4-1}

\usepackage{graphicx}
\usepackage{dcolumn}
\usepackage{bm}
\usepackage{lipsum}
\usepackage[most]{tcolorbox}
\usepackage[section]{placeins}
\usepackage{float}
\usepackage{mwe}

\begin{document}

\title{Hybrid functional calculations of electronic structure and carrier densities in rare-earth monopnictides}

\author{Shoaib Khalid, Abhishek Sharan}
\affiliation{Department of Physics and Astronomy, University of Delaware, Newark, DE 19716, USA}
\author{Anderson Janotti}
\email{janotti@udel.edu}
\affiliation{Department of Material Science and Engineering, University of Delaware, Newark, DE 19716, USA}

\begin{abstract}
The structural parameters and electronic structure of rare-earth pnictides are calculated using density functional theory (DFT) with the Heyd, Scuseria, and Ernzerhof (HSE06) screened hybrid functional.  We focus on RE-V compounds, with RE=La, Gd, Er, and Lu, and V=As, Sb, and Bi, and analyze the effects of spin-orbit coupling and treating the RE 4$f$ electrons as valence electrons in the projector augmented wave approach. The results of HSE06 calculations are compared with DFT within the generalized gradient approximation (DFT-GGA) and other previous calculations. We find that all these RE-V compounds are semimetals with electron pockets at the $X$ point and hole pockets at $\Gamma$.  Whereas in DFT-GGA the carrier density is significantly overestimated, the computed carrier densities using HSE06 is in good agreement with the available experimental data. 

\end{abstract}


\maketitle

\section{Introduction} \label{sec:intro}

Rare-earth monopnictides (RE-V) display interesting electronic, magnetic, optical and magneto-optical properties, with applications including thermoelectrics \cite{zide2010high}, tunnel junctions \cite{gossard2007enhanced}, photoconductive switches, and terahertz detectors\cite{middendorf2012thz}.
RE-V have the rocksalt structure, which is compatible with the zinc-blende structure of III-V semiconductors. It has been demonstrated that RE-V compounds can be epitaxially grown on III-V semiconductors\cite{palmstrom1992,klenov2005interface,shoaib2019}, to which RE-V have been explored as the ultimate ohmic contacts with high structural quality due to the small lattice mismatch \cite{hanson2007controlling}. For instance, ErAs and TbAs have lattice parameters very close to those of GaAs and InGaAs alloys \cite{zide2016,zide2017,Zhang2019}, respectively. 
Some of the RE-V compounds have been investigated due to their non-trivial topological band structures\cite{niu2016presence,lou2017evidence,khalid2018topological}, and have shown effects of extreme magnetoresistance \cite{zeng2016compensated,guo2016charge} and superconductivity \cite{tafti2017tuning} at low temperatures.

The electronic and magnetic properties of RE-V have long been studied both experimentally\cite{tafti2017tuning,allen1989eras,abdusalyamova2002investigation} and by theory \cite{child1963neutron,petit2016rare,sclar1962energy,khalid2018topological}. There have been notable contradictions in the experimental characterization of the electronic properties of these materials. Early  measurements of electrical resistivity have shown metallic as well as semiconducting behavior \cite{Andersen1980}, while optical measurements have shown signs of a semiconducting band gap \cite{Eyring1999}.
Previous theoretical work were limited to semi-classical treatment of the crystalline field\cite{trammell1963magnetic}, or simplified models that account for the $p$-$f$ mixing \cite{Takahashi1985}, $d$-$f$ coulomb interaction \cite{narita1985magnetic}, and an effective point-charge model for the crystalline field \cite{birgeneau1973crystal}. These models could clarify only few specific properties of some RE-V compounds, yet general agreement with experiments for the complete series was not satisfactory \cite{hasegawa1977energy}. 
Using an augmented plane wave method with the Slater X$_\alpha$ exchange potential \cite{slater1951simplification} and treating the 4$f$ electrons as core electrons, Hasegawa and Yanase \cite{hasegawa1977energy} claimed that GdN is a semiconductor with a band gap of 1 eV and all the other Gd monopnictides are semimetallic.

A remarkable feature of RE-V compounds is the presence of occupied 4$f$ electronic states near to or resonant in the valence band. As the number of 4$f$ electrons increases from La (no $f$ electrons) to Lu (fully occupied 4$f$ shell), the RE-V series displays a variety of magnetic and electronic effects. The coexistence of partially filled 4$f$ shell along with itinerant $p$ and $d$ charge carriers has been quite challenging to an accurate description of the electronic structure of RE-V compounds.
Petukhov {\em et al.} \cite{petukhov1994electronic} performed first-principles calculations using the linear-muffin-tin-orbital (LMTO) method within local spin-density approximation for ErAs and Er$_{1-x}$Sc$_x$As,  treating the Er 4$f$ electrons as core-like electrons. They found cyclotron masses in good agreement with experimental data \cite{allen1989eras}, however, the Fermi surface dimensions were significantly overestimated.  Subsequently, Petukhov {\em et al.} \cite{petukhov1996electronic} studied the electronic properties of GdX and ErX (X = N,P,As), performing test calculations for ErAs and GdAs with 4$f$ electrons in the valence and in the core. 
They reported that treating the 4$f$ electrons in the valence leads to strong perturbation of the bands near the Fermi level, and incorrectly predicts that
these compounds are not semi-metals. 
They also claimed that GdN is metallic for one spin channel and semiconducting in the other. Later studies treated 4$f$ as core electrons \cite{hasegawa1977energy,brooks19913d} while some others highlighted the need of including the $f$ electrons in the valence \cite{temmerman1990band,heinemann1994magnetic}. 

More recently, a combination of DFT and dynamical mean-field theory (DMFT) calculations indicate the importance of including the 4$f$ electrons in the valence to correctly describe the dimensions of the Fermi surface pockets, carrier
concentration, and Shubnikov-de Haas (SdH) oscillation frequencies \cite{pourovskii2009role}.
These calculations are rather computationally expensive, and finding other more computationally affordable methods that correctly describe the effects 4$f$ electrons on the electronic structure of RE-V compounds is highly desirable. Here we studied
the electronic structure of RE-V compounds using the screened hybrid functional of Heyd, Scuseria, and Ernzerhof (HSE06), focusing on the effects of spin-orbit coupling and treating the 4$f$ as valence electrons.  
 
\section{Computational Method} \label{sec:Com}

We use density functional theory (DFT) \cite{hohenberg1964inhomogeneous,kohn1965self} with the screened hybrid functional of Heyd-Scuseria-Ernzerhof (HSE06) \cite{heyd2003hybrid,HSE} as implemented in the VASP code \cite{kresse1993ab,kresse1994ab}. %
In HSE06, the exchange potential is divided into short range and long range parts, and the Hartree-Fock exchange is mixed with the exchange of the generalized gradient approximation (GGA) \cite{perdew1996generalized}, with a mixing parameter $\alpha$ of 0.25, only in the short range part. The long range part is described by the GGA exchange potential. 
The interaction between valence electrons and the ion cores
is described by the projector augmented wave (PAW) method\cite{blochl1994projector}. The PAW potentials for the pnictides As, Sb and Bi all contain 5 valence electrons each, i.e  As: 4s\textsuperscript{2}4p\textsuperscript{3}, Sb: 5s\textsuperscript{2}5p\textsuperscript{3}, Bi: 6s\textsuperscript{2}6p\textsuperscript{3}. For the rare-earth elements La, Gd, Er and Lu, the PAW potentials with 4$f$ electrons in the core contain 11, 9, 9 and 9 valence electrons, respectively, i.e La: 5s\textsuperscript{2}5p\textsuperscript{6}6s\textsuperscript{2}5d\textsuperscript{1}, Gd: 5p\textsuperscript{6}6s\textsuperscript{2}5d\textsuperscript{1}, Er: 5p\textsuperscript{6}6s\textsuperscript{2}5d\textsuperscript{1}, Lu: 5p\textsuperscript{6}6s\textsuperscript{2}5d\textsuperscript{1}. 
In the calculations where the 4$f$ are treated as valence electrons, we used PAW potentials
with the following valence configurations:
La: 5s\textsuperscript{2}5p\textsuperscript{6}6s\textsuperscript{2}5d\textsuperscript{1}4f\textsuperscript{0}, Gd: 5p\textsuperscript{6}6s\textsuperscript{2}5d\textsuperscript{1}4f\textsuperscript{7}, Er:5p\textsuperscript{6}6s\textsuperscript{2}5d\textsuperscript{1}4f\textsuperscript{12}, Lu: 5p\textsuperscript{6}6s\textsuperscript{2}5d\textsuperscript{1}4f\textsuperscript{14}. 
We note that the calculations including the 4$f$ as valence electrons were only
performed using HSE06, since GGA does not correctly describe the states originating from 4$f$ orbitals, placing them at the Fermi level when the $f$ shell is partially filled \cite{petukhov1994electronic}.

The rocksalt crystal structure of the RE-V compounds has two atoms in the primitive cell, with the rare-earth atom at (0,0,0) and the  pnictide at (0.5,0.5,0.5). All the calculations were performed using a 400 eV cutoff for the plane-wave basis set, and 12$\times$12$\times$12 $\Gamma$-centered mesh of $k$ points for the integrations over the Brillouin zone.

For the compounds with Gd, Er, and Lu, we only considered the ferromagnetic ordering using the 2-atom primitive cells when including the 4$f$ electrons in the valence, despite these compounds are known to be antiferromagnetic at low temperatures \cite{li1997magnetic,narita1985magnetic,allen1990magneto}.  This facilitates the comparison of the band structures of the different compounds, without having to deal with folded bands in the smaller Brillouin zone of the larger cell size required to describe the anti ferromagnetic ordering. We expect the magnetic ordering to not change our results and conclusions. 

\section{Results and Discussion}\label{sec:result}

\subsection{Lattice parameters of RE-V}

The calculated  lattice parameters of RE-V compounds using HSE06 along with the results from GGA calculations and experimental values are listed in Table~\ref{lattice}. The results shown were obtained by treating the 4$f$ electrons as core and valence electrons. We note that the HSE06 results are systematically closer to the experimental values than the results from the DFT-GGA, which typically slightly overestimates lattice parameters.

\begin{table}
\caption{Calculated equilibrium lattice parameters $a$ for the RE-V compounds using the HSE06 hybrid functional. The results obtained using DFT-GGA and the experimental values are also shown for comparison \cite{petukhov1994electronic,abdusalyamova2002investigation,mullen1974magnetic,li1997magnetic,hasegawa1977energy,chua1974simple,shirotani2003pressure}. For the La-V compounds, the 4$f$ shell is empty so the results using HSE06 with the 4$f$ in the core or valence are the same. For Gd, Er, and Lu, treating the 4$f$ as core electrons or as valence electrons give slightly different lattice parameters.}
\centering
\begin{tabular}{l c c c c}
\hline\hline
Material &DFT-GGA         & HSE06                    &    HSE06  &Exp.\\
         & (4$f$ in the core) & (4$f$ in the core)   & (4$f$ in the valence) &  \\
         & $a$ (\AA)  & $a$ (\AA) & $a$ (\AA)  & $a$ (\AA)\\[0.5ex] 
\hline 
LaAs & 6.187 & 6.173 & 6.173 & 6.137 \\
LaSb & 6.540 & 6.514 & 6.514 & 6.488 \\
LaBi & 6.654 & 6.625 & 6.625 & 6.578 \\
GdAs & 5.879 & 5.838 & 5.882 & 5.854 \\
GdSb & 6.247 & 6.192 & 6.245 & 6.217 \\
GdBi & 6.373 & 6.314 & 6.368 & 6.295 \\
ErAs & 5.769 & 5.737 & 5.766 & 5.732 \\
ErSb & 6.148 & 6.105 & 6.160 & 6.106 \\
ErBi & 6.281 & 6.233 & 6.269 & 6.206 \\
LuAs & 5.701 & 5.670 & 5.697 & 5.679 \\
LuSb & 6.091 & 6.056 & 6.081 & 6.055 \\
LuBi & 6.231 & 6.185 & 6.220 & 6.159 \\[0.5ex]
\hline\hline 
\end{tabular}
\label{lattice} 
\end{table}

We note from Table~\ref{lattice} that for a given rare-earth, the lattice parameter $a$ increases going from As, Sb, to Bi, i.e, as the atomic size of the pnictide atom increases. On the other hand, the lattice parameter decreases by going from La, Gd, Er, to Lu, due to the lanthanide contraction effect  on the atomic size \cite{cotton1988advanced}.


\subsection{Electronic band structures of the RE-V}

\begin{figure*}
\centering
\includegraphics[width=6.5in]{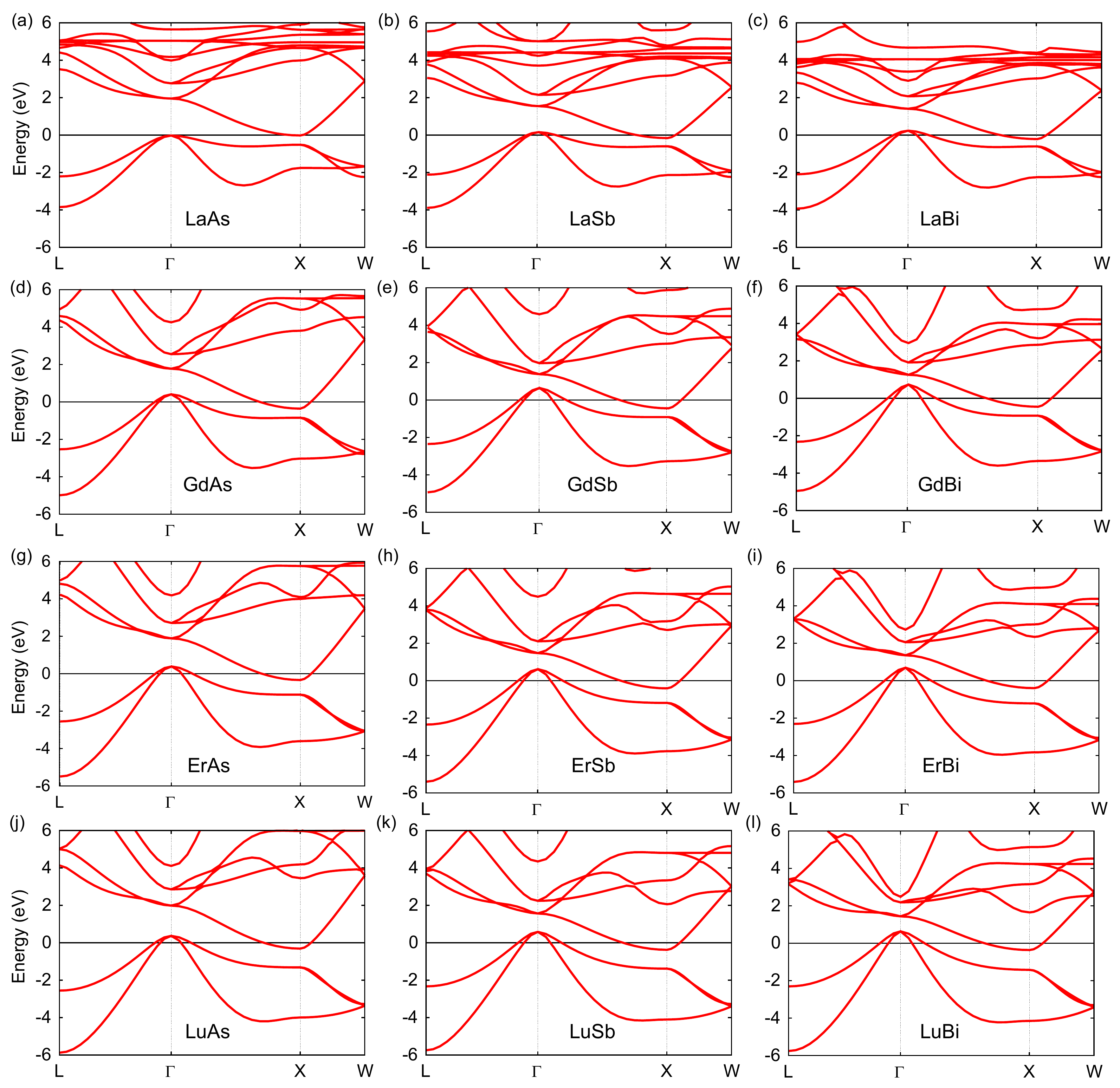}
\caption{Calculated electronic band structure of (a)LaAs, (b) LaSb, (c) LaBi, (d) GdAs, (e) GdSb, (f) GdBi, (g) ErAs, (h) ErSb, (i) ErBi, (j) LuAs, (k) LuSb, and (l) LuBi using HSE06 without spin-orbit coupling and treating the 4$f$ electrons as core electrons. The Fermi level is set to zero.}
\label{fig1}
\end{figure*}

The analysis of the electronic structure of the RE-V compounds starts with the results using HSE06 without spin-orbit coupling and without including the 4$f$ electrons in the valence, shown in Figure \ref{fig1}. In this approximation, all the RE-V studied are semi-metallic except for LaAs which shows a very small band gap of 5 meV. All RE-V compounds have hole pockets at $\Gamma$ point and electron pockets at the $X$ point. The hole pocket bands are composed mostly of pnictide $p$ orbitals while the electron pocket bands has derived mostly from the rare-earth $d$ orbitals. For a given rare-earth we note that the size of the hole pocket increases going from As to Bi. This trend is explained by the relative energy of the valence $p$ orbitals of the pnictide atoms, which increases from As to Bi \cite{harrison2012electronic}. The gap at $X$ point opens up as we go from La to Lu largely due to the dispersion of the pnictide $p$ band that increases as the lattice parameter decreases due to the lanthanide contraction effect. 

\subsubsection{Effects of spin-orbit coupling on the electronic structure of RE-V}

\begin{figure*}
   \centering
   \includegraphics[width=6.5in]{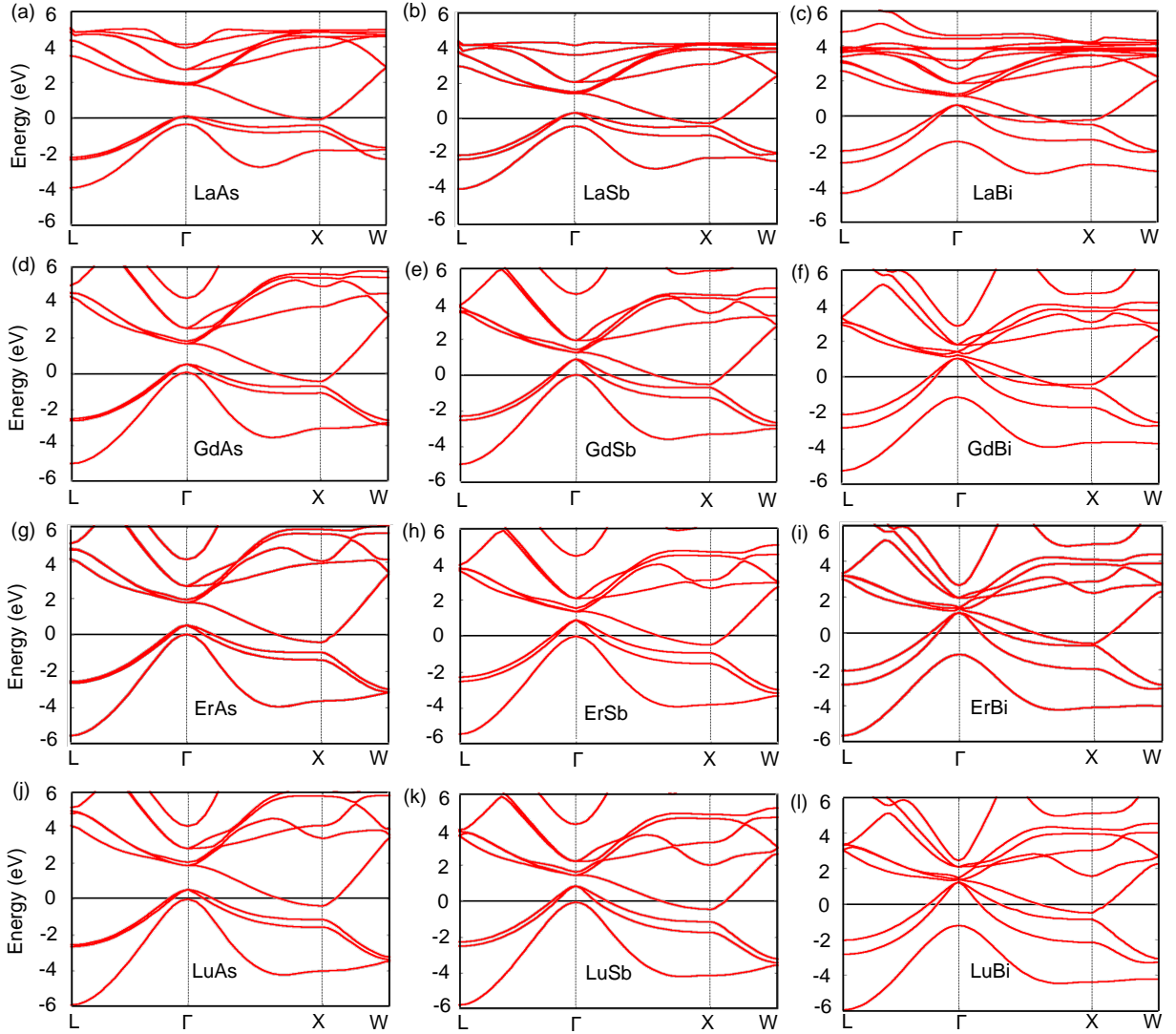}
   \caption{Calculated electronic band structure of (a)LaAs, (b) LaSb, (c) LaBi, (d) GdAs, (e) GdSb, (f) GdBi, (g) ErAs, (h) ErSb, (i) ErBi, (j) LuAs, (k) LuSb, and (l) LuBi using HSE06 with spin-orbit coupling and treating the 4$f$ electrons as core electrons. The Fermi level is set to zero.}
   \label{fig2}
\end{figure*}

The effects of spin-orbit coupling are very significant for the RE-V compounds, which are composed of heavy elements and the bands of interest, situated near the Fermi level, are derived from pnictide $p$ and lanthanide $d$ orbitals.
Previous first-principles calculations have already pointed out the importance of spin-orbit coupling in the description of the electronic structure of RE-Vs \cite{duan2004hybridization,leuenberger2005gdn}. The three fold degenerate pnicitide $p$ band at $\Gamma$ splits into 2+1 bands due to spin-orbit coupling. Also the splitting increases from As to Bi as the atomic number increases.
The spin-orbit coupling also causes a splitting of the highest occupied pnictide $p$ band at the $X$ point, and this splitting also increases going from As to Bi. Interestingly, as shown in Figure~\ref{fig2} for the case of Er-V, we find that in ErBi, the Bi 6$p$ band touches the Er 5$d$ band at the $X$ point, making it a topological semimetal. This crossing is affected by the presence of 4$f$ bands, which is discussed below.

\subsection{Effects of including 4$f$ as valence electrons on the electronic structure of RE-V}

\begin{figure}
\begin{center}
   \includegraphics[width=6.5in]{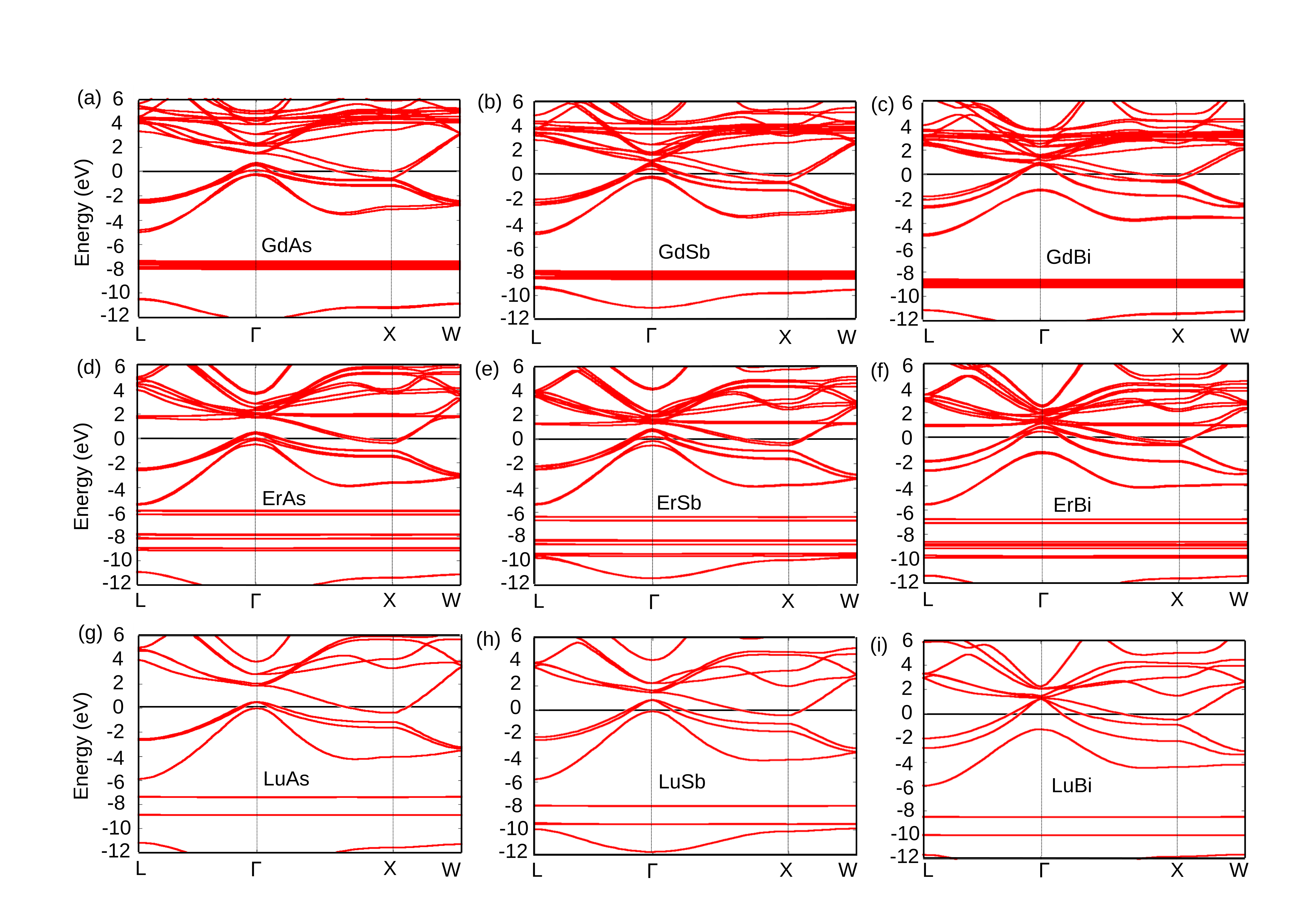}
\end{center}
  \caption{Calculated electronic band structure of (a) GdAs, (b) GdSb, (c) GdBi, (d) ErAs, (e) ErSb, (f) ErBi, (g) LuAs, (h) LuSb, and (i)  LuBi with $f$ states as bands using HSE06.}
  \label{fig3}
\end{figure}

In most of the previous calculations of RE-V compounds using DFT-LSDA or GGA the 4$f$ electrons were taken as core electrons.  Treating the 4$f$ as valence electrons in these approximations would lead to incorrect description of the bands near the Fermi level \cite{petukhov1996electronic}. To overcome this problem, an extra Coulomb interaction is often added to the 4$f$ orbitals as in the DFT+U method, with $U$ typically used as an adjusting parameter. This added electron-electron repulsion term splits the occupied and unoccupied 4$f$ bands, pushing them out of the Fermi level region. Here, instead, we show that 
the HSE06 hybrid functional greatly improves the description of the electronic structure of RE-V compounds, including the effects of 4$f$ electrons being treated self-consistently as valence electrons. The results for GdAs, ErAs, and LuAs are shown in Figure~\ref{fig3}.

For GdAs, the 4$f$ shell is half filled, resulting in flat bands occupied well below the Fermi level, whereas the empty 4$f$ bands lie well above the Fermi level. The occupied 4$f$ bands are located around 8 eV below the Fermi level and the unoccupied 4$f$ states lie around 4 eV above the Fermi level, in qualitative agreement  with the results of Petukhov {\em et al.} \cite{petukhov1996electronic}. In ErAs, more than half of 4$f$ orbital is filled and the filled bands lie between 5 eV and 9 eV below the Fermi level, while the unoccupied 4$f$ bands are at 2 eV above the Fermi level. The splitting and position of the 4$f$ bands in ErAs are in good agreement with recent DMFT calculations \cite{pourovskii2009role}. In the case of LuAs, the 4$f$ orbital is completely filled, and the 4$f$ bands lie in the range of 7 eV  and 9 eV below Fermi level.

\subsection{Carrier concentrations in RE-V}

The overlap in energy of the electron and hole pockets determine the free carrier concentration in these materials. The volume of the Fermi surface for ErAs calculated using LSDA \cite{petukhov1994electronic} is almost three times larger than the experimental value \cite{bogaerts1996experimental}. The reason behind this disagreement is the fact that LSDA overestimates the band overlap between As $p$ and Er $d$ bands, overestimating the dimensions of the electron and hole pockets and resulting in higher carrier concentrations than in experiment.  Here we computed the carrier concentration using different approximations, i.e., GGA, HSE06, including SOC, and treating the 4$f$ as core electrons. The carrier concentrations are calculated using the wannier90\cite{MOSTOFI20142309} and SuperCell K-space Extremal Area Finder(SKEAF) codes \cite{julian2012numerical}. The results are shown in Table \ref{table:concentration}. 
Only the results including SOC should be compared to the experimental data.

\begin{table}
\caption{Calculated carrier concentration $n$ for the RE-V compounds using the HSE06 hybrid functional, including spin-orbit coupling (SOC) and treating the $f$ electrons as core (DFT-GGA) or valence (HSE06) electrons  Experimental results are also listed for comparison.\cite{allen1989eras,allen1990magneto,bogaerts1996experimental,bogaerts1993experimental,allen1991band,yang2017extreme,tafti2016resistivity,kasuya1996normal,sun2016large,li1996electrical,zogal2014electron,pavlosiuk2017fermi,pavlosiuk2018magnetoresistance} }

\centering
\begin{tabular}{l c c c}
\hline\hline
Material &DFT-GGA &HSE06  &Exp.\\
 & $n$($10^{20}$ cm$^{-3}$)  & $n$($10^{20}$ cm$^{-3}$) & $n$($10^{20}$ cm$^{-3}$)\\[0.5ex] 
\hline 
LaAs & 0.94 & 0.25 & 0.46\cite{yang2017extreme} \\
LaSb & 2.28 & 1.44 & 1.10\cite{tafti2016resistivity} \\
LaBi & 4.04 & 3.72 & 3.78 \cite{kasuya1996normal,sun2016large} \\
GdAs & 5.13 & 3.00 & 2.10\cite{li1996electrical} \\
GdSb & 6.55 & 4.39 & 4.20\cite{li1996electrical} \\
GdBi & 7.18 & 6.09 & -\\
ErAs & 5.59   & 3.3   & 1.8\cite{allen1989eras,allen1990magneto},4.69\cite{bogaerts1996experimental,bogaerts1993experimental},3.3\cite{allen1989eras,allen1990magneto,allen1991band} \\
ErSb & 6.68 & 4.53 & -\\
ErBi & 8.06 & 6.88 & -\\
LuAs & 5.86 & 2.63 & 1.52\cite{zogal2014electron} \\
LuSb & 6.57 & 4.35 & 4.35,5.07\cite{pavlosiuk2017fermi} \\
LuBi & 8.24 & 6.44 & 6.61,6.99\cite{pavlosiuk2018magnetoresistance} \\[0.5ex]
\hline\hline 
\end{tabular}
\label{table:concentration} 
\end{table}

Due to the compatibility with conventional III-V semiconductors, we expect that the later can serve as substrates for epitaxial growth of RE-V, as demonstrated in the case of (In)GaAs/ErAs \cite{klenov2005interface,zide2016,zide2017} and GaSb/LuSb \cite{shoaib2019}. In this context, it is important to know the workfunction of these materials to understand the formation of Schottky barriers and any possible charge transfer across the III-V/RE-V interfaces.  So we calculated the band alignment at the ErAs/GaAs and LuSb/GaSb, which are two systems of current interest, and display small lattice mismatches \cite{delaney2010,shoaib2019}.

\begin{figure}
\includegraphics[width=13cm]{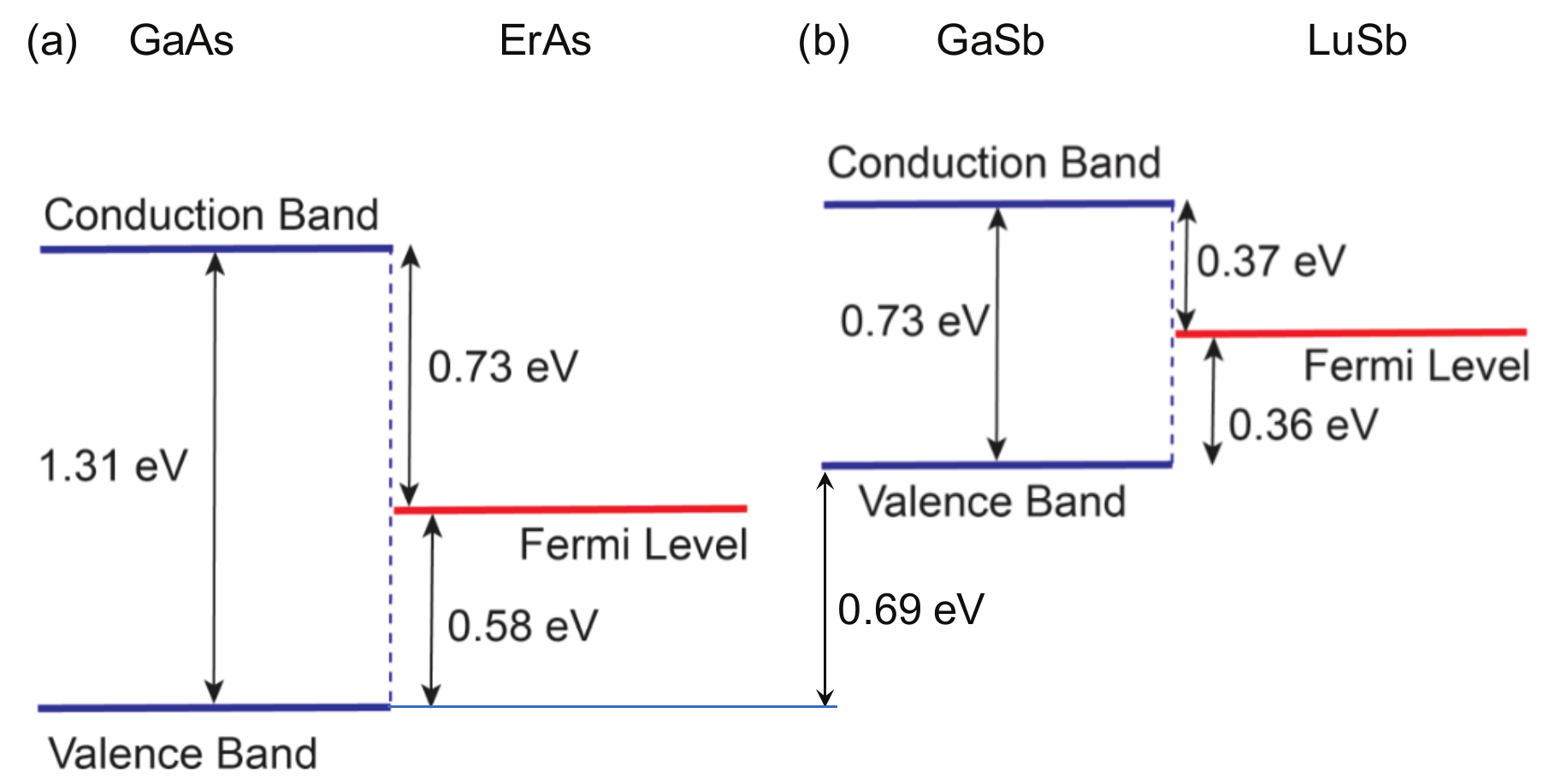}
\caption{Calculated band alignment at the ErAs/GaAs and LuSb/GaSb interfaces.The band alignment between GaAs and GaSb was taken from the literature \cite{vandewalle2003}.
}
\label{fig4}
\end{figure}

The band alignments between the Fermi level in the semimetal and the band edges in the semiconductor are calculated as follows \cite{delaney2010}. First we calculated the Fermi level position in the semimetal and the band edges in the semiconductor with respect to the respective averaged electrostatic potential through bulk calculations using their primitive cells.  Then we aligned the averaged electrostatic potential in the two materials by performing an interface calculation using a superlattice geometry with two equivalent interfaces, as described previously.  The superlattice consisted of 9 monolayers of each material, along the [110] direction in order to minimize the effect of charge transfer across the interface. In order to remove the effects of strain due to the small lattice parameters, we used the in-plane lattice parameter of the semiconductor and adjusted the out-of-plane lattice parameter of the semimetal to give its equilibrium volume. In this way, the calculated band alignments should be regarded as natural band alignments. in any case, effects of small strain on the absolute position of the Fermi level in the semimetal were shown to be negligible \cite{delaney2010}.

For the superlattice we used a 6$\times$6$\times$1 Gamma-centered mesh of special $k$ points for the integration over the Brillouin zone.  We note than for GaAs our calculated band gap using the HSE06 hybrid functional is 0.11 eV lower than the experimental value of 1.42 eV, while for GaSb our calculated band gap is 0.06 eV larger than the experimental value of 0.67 eV. The results for the band alignments are shown in Fig.~\ref{fig4}.

We find that the the Fermi level of the semimetal lies within the band gap of the semiconductor in the cases of ErAs/GaAs and LuSb/GaSb. The band alignment between GaAs and GaSb was taken from the literature. In the case of ErAs/GaAs we find that the Fermi level in ErAs is 0.58 eV above the valence band of GaAs, and in the case of LuSb/GaSb the Fermi level of LuSb is 0.36 eV above the valence band of GaSb. From the transitivity rule, we obtain the Fermi level of LuSb 0.47 eV higher than that of ErAs, consistent with the Sb 5$p$ orbitals being higher in energy than the As 4$p$ orbitals. Note that our results for position of the Fermi level in ErAs with respect to the valence band edge in GaAs is higher than previous uncorrected DFT value by 0.3 eV \cite{delaney2010}, yet only $\sim$0.1 eV lower than the experimental value \cite{palmstrom1992}.

\section{Summary}\label{sec:Summary}

We performed HSE06 hybrid functional calculations for the electronic structure of RE-V compounds, where RE= La, Gd, Er, Lu and V = As, Sb, Bi. The HSE06 gives equilibrium lattice parameters in very close agreement with experimental data. All the studied compounds are semi-metals with the size of hole pocket increasing from As to Bi, whereas the gap at $X$ point increases with increase in atomic number of the rare-earth element. We show that using HSE06 gives a good description of the electronic structure of RE-V including the 4$f$ electrons explicitly as valence electrons. In particular, we find good agreement with DMFT calculations for ErAs. We also show that HSE06 gives hole and electron concentrations that are closer to the observed values, correcting the overestimation of the electron and hole pockets overlap in the DFT-GGA calculations. Our results for the band alignment at the ErAs/GaAs interface is in good agreement with experimental data, and it is a significant improvement over previously reported DFT calculations within local spin density approximation. 

\section*{Acknowledgements}
We acknowledge fruitful discussions with C. J. Palmstr\o{}m and S. Chatterjee. This work was supported by the U.S. Department of Energy Basic Energy Science under Award
No. DE-SC0014388 and the National Energy Research Scientific Computing Center (NERSC), a U.S. Department of Energy Office of Science User Facility operated under Contract No. DE-AC02-05CH11231.


\begin{thebibliography}{67}%
\makeatletter
\providecommand \@ifxundefined [1]{%
 \@ifx{#1\undefined}
}%
\providecommand \@ifnum [1]{%
 \ifnum #1\expandafter \@firstoftwo
 \else \expandafter \@secondoftwo
 \fi
}%
\providecommand \@ifx [1]{%
 \ifx #1\expandafter \@firstoftwo
 \else \expandafter \@secondoftwo
 \fi
}%
\providecommand \natexlab [1]{#1}%
\providecommand \enquote  [1]{``#1''}%
\providecommand \bibnamefont  [1]{#1}%
\providecommand \bibfnamefont [1]{#1}%
\providecommand \citenamefont [1]{#1}%
\providecommand \href@noop [0]{\@secondoftwo}%
\providecommand \href [0]{\begingroup \@sanitize@url \@href}%
\providecommand \@href[1]{\@@startlink{#1}\@@href}%
\providecommand \@@href[1]{\endgroup#1\@@endlink}%
\providecommand \@sanitize@url [0]{\catcode `\\12\catcode `\$12\catcode
  `\&12\catcode `\#12\catcode `\^12\catcode `\_12\catcode `\%12\relax}%
\providecommand \@@startlink[1]{}%
\providecommand \@@endlink[0]{}%
\providecommand \url  [0]{\begingroup\@sanitize@url \@url }%
\providecommand \@url [1]{\endgroup\@href {#1}{\urlprefix }}%
\providecommand \urlprefix  [0]{URL }%
\providecommand \Eprint [0]{\href }%
\providecommand \doibase [0]{http://dx.doi.org/}%
\providecommand \selectlanguage [0]{\@gobble}%
\providecommand \bibinfo  [0]{\@secondoftwo}%
\providecommand \bibfield  [0]{\@secondoftwo}%
\providecommand \translation [1]{[#1]}%
\providecommand \BibitemOpen [0]{}%
\providecommand \bibitemStop [0]{}%
\providecommand \bibitemNoStop [0]{.\EOS\space}%
\providecommand \EOS [0]{\spacefactor3000\relax}%
\providecommand \BibitemShut  [1]{\csname bibitem#1\endcsname}%
\let\auto@bib@innerbib\@empty
\bibitem [{\citenamefont {Zide}\ \emph {et~al.}(2010)\citenamefont {Zide},
  \citenamefont {Bahk}, \citenamefont {Singh}, \citenamefont {Zebarjadi},
  \citenamefont {Zeng}, \citenamefont {Lu}, \citenamefont {Feser},
  \citenamefont {Xu}, \citenamefont {Singer}, \citenamefont {Bian},
  \citenamefont {Majumdar}, \citenamefont {Bowers}, \citenamefont {Shakouri},\
  and\ \citenamefont {Gossard}}]{zide2010high}%
  \BibitemOpen
  \bibfield  {author} {\bibinfo {author} {\bibfnamefont {J.~M.~O.}\
  \bibnamefont {Zide}}, \bibinfo {author} {\bibfnamefont {J.-H.}\ \bibnamefont
  {Bahk}}, \bibinfo {author} {\bibfnamefont {R.}~\bibnamefont {Singh}},
  \bibinfo {author} {\bibfnamefont {M.}~\bibnamefont {Zebarjadi}}, \bibinfo
  {author} {\bibfnamefont {G.}~\bibnamefont {Zeng}}, \bibinfo {author}
  {\bibfnamefont {H.}~\bibnamefont {Lu}}, \bibinfo {author} {\bibfnamefont
  {J.~P.}\ \bibnamefont {Feser}}, \bibinfo {author} {\bibfnamefont
  {D.}~\bibnamefont {Xu}}, \bibinfo {author} {\bibfnamefont {S.~L.}\
  \bibnamefont {Singer}}, \bibinfo {author} {\bibfnamefont {Z.~X.}\
  \bibnamefont {Bian}}, \bibinfo {author} {\bibfnamefont {A.}~\bibnamefont
  {Majumdar}}, \bibinfo {author} {\bibfnamefont {J.~E.}\ \bibnamefont
  {Bowers}}, \bibinfo {author} {\bibfnamefont {A.}~\bibnamefont {Shakouri}}, \
  and\ \bibinfo {author} {\bibfnamefont {A.~C.}\ \bibnamefont {Gossard}},\
  }\href@noop {} {\bibfield  {journal} {\bibinfo  {journal} {J. Appl. Phys.}\
  }\textbf {\bibinfo {volume} {108}},\ \bibinfo {pages} {123702} (\bibinfo
  {year} {2010})}\BibitemShut {NoStop}%
\bibitem [{\citenamefont {Gossard}\ \emph {et~al.}(2007)\citenamefont
  {Gossard}, \citenamefont {Zide},\ and\ \citenamefont
  {Zimmerman}}]{gossard2007enhanced}%
  \BibitemOpen
  \bibfield  {author} {\bibinfo {author} {\bibfnamefont {A.}~\bibnamefont
  {Gossard}}, \bibinfo {author} {\bibfnamefont {J.}~\bibnamefont {Zide}}, \
  and\ \bibinfo {author} {\bibfnamefont {J.}~\bibnamefont {Zimmerman}},\
  }\href@noop {} {\bibfield  {journal} {\bibinfo  {journal} {Google Patents}\ }
  (\bibinfo {year} {2007})},\ \bibinfo {note} {{US Patent App.
  11/675,269}}\BibitemShut {NoStop}%
\bibitem [{\citenamefont {Middendorf}\ and\ \citenamefont
  {Brown}(2012)}]{middendorf2012thz}%
  \BibitemOpen
  \bibfield  {author} {\bibinfo {author} {\bibfnamefont {J.~R.}\ \bibnamefont
  {Middendorf}}\ and\ \bibinfo {author} {\bibfnamefont {E.~R.}\ \bibnamefont
  {Brown}},\ }\href@noop {} {\bibfield  {journal} {\bibinfo  {journal} {Opt.
  Express}\ }\textbf {\bibinfo {volume} {20}},\ \bibinfo {pages} {16504}
  (\bibinfo {year} {2012})}\BibitemShut {NoStop}%
\bibitem [{\citenamefont {Palmstr\o{}m}\ \emph {et~al.}(1992)\citenamefont
  {Palmstr\o{}m}, \citenamefont {Cheeks}, \citenamefont {Gilchrist},
  \citenamefont {Zhu}, \citenamefont {Carter}, \citenamefont {Wilkens},\ and\
  \citenamefont {Martin}}]{palmstrom1992}%
  \BibitemOpen
  \bibfield  {author} {\bibinfo {author} {\bibfnamefont {C.~J.}\ \bibnamefont
  {Palmstr\o{}m}}, \bibinfo {author} {\bibfnamefont {T.~L.}\ \bibnamefont
  {Cheeks}}, \bibinfo {author} {\bibfnamefont {H.~L.}\ \bibnamefont
  {Gilchrist}}, \bibinfo {author} {\bibfnamefont {J.~G.}\ \bibnamefont {Zhu}},
  \bibinfo {author} {\bibfnamefont {C.~B.}\ \bibnamefont {Carter}}, \bibinfo
  {author} {\bibfnamefont {B.~J.}\ \bibnamefont {Wilkens}}, \ and\ \bibinfo
  {author} {\bibfnamefont {R.}~\bibnamefont {Martin}},\ }\href@noop {}
  {\bibfield  {journal} {\bibinfo  {journal} {â€ŽJ. Vac. Sci. Technol A}\
  }\textbf {\bibinfo {volume} {10}},\ \bibinfo {pages} {1946} (\bibinfo {year}
  {1992})}\BibitemShut {NoStop}%
\bibitem [{\citenamefont {Klenov}\ \emph {et~al.}(2005)\citenamefont {Klenov},
  \citenamefont {Zide}, \citenamefont {Zimmerman}, \citenamefont {Gossard},\
  and\ \citenamefont {Stemmer}}]{klenov2005interface}%
  \BibitemOpen
  \bibfield  {author} {\bibinfo {author} {\bibfnamefont {D.~O.}\ \bibnamefont
  {Klenov}}, \bibinfo {author} {\bibfnamefont {J.~M.}\ \bibnamefont {Zide}},
  \bibinfo {author} {\bibfnamefont {J.~D.}\ \bibnamefont {Zimmerman}}, \bibinfo
  {author} {\bibfnamefont {A.~C.}\ \bibnamefont {Gossard}}, \ and\ \bibinfo
  {author} {\bibfnamefont {S.}~\bibnamefont {Stemmer}},\ }\href@noop {}
  {\bibfield  {journal} {\bibinfo  {journal} {Appl. Phys. Lett.}\ }\textbf
  {\bibinfo {volume} {86}},\ \bibinfo {pages} {241901} (\bibinfo {year}
  {2005})}\BibitemShut {NoStop}%
\bibitem [{\citenamefont {Chatterjee}\ \emph {et~al.}(2019)\citenamefont
  {Chatterjee}, \citenamefont {Khalid}, \citenamefont {Inbar}, \citenamefont
  {Goswami}, \citenamefont {de~Lima}, \citenamefont {Sharan}, \citenamefont
  {Sabino}, \citenamefont {Brown-Heft}, \citenamefont {Chang}, \citenamefont
  {Fedorov}, \citenamefont {Read}, \citenamefont {Janotti},\ and\ \citenamefont
  {Palmstr\o{}m}}]{shoaib2019}%
  \BibitemOpen
  \bibfield  {author} {\bibinfo {author} {\bibfnamefont {S.}~\bibnamefont
  {Chatterjee}}, \bibinfo {author} {\bibfnamefont {S.}~\bibnamefont {Khalid}},
  \bibinfo {author} {\bibfnamefont {H.~S.}\ \bibnamefont {Inbar}}, \bibinfo
  {author} {\bibfnamefont {A.}~\bibnamefont {Goswami}}, \bibinfo {author}
  {\bibfnamefont {F.~C.}\ \bibnamefont {de~Lima}}, \bibinfo {author}
  {\bibfnamefont {A.}~\bibnamefont {Sharan}}, \bibinfo {author} {\bibfnamefont
  {F.~P.}\ \bibnamefont {Sabino}}, \bibinfo {author} {\bibfnamefont {T.~L.}\
  \bibnamefont {Brown-Heft}}, \bibinfo {author} {\bibfnamefont {Y.-H.}\
  \bibnamefont {Chang}}, \bibinfo {author} {\bibfnamefont {A.~V.}\ \bibnamefont
  {Fedorov}}, \bibinfo {author} {\bibfnamefont {D.}~\bibnamefont {Read}},
  \bibinfo {author} {\bibfnamefont {A.}~\bibnamefont {Janotti}}, \ and\
  \bibinfo {author} {\bibfnamefont {C.~J.}\ \bibnamefont {Palmstr\o{}m}},\
  }\href@noop {} {\bibfield  {journal} {\bibinfo  {journal} {Phys. Rev. B}\
  }\textbf {\bibinfo {volume} {99}},\ \bibinfo {pages} {125134} (\bibinfo
  {year} {2019})}\BibitemShut {NoStop}%
\bibitem [{\citenamefont {Hanson}\ \emph {et~al.}(2007)\citenamefont {Hanson},
  \citenamefont {Bank}, \citenamefont {Zide}, \citenamefont {Zimmerman},\ and\
  \citenamefont {Gossard}}]{hanson2007controlling}%
  \BibitemOpen
  \bibfield  {author} {\bibinfo {author} {\bibfnamefont {M.~P.}\ \bibnamefont
  {Hanson}}, \bibinfo {author} {\bibfnamefont {S.~R.}\ \bibnamefont {Bank}},
  \bibinfo {author} {\bibfnamefont {J.~M.~O.}\ \bibnamefont {Zide}}, \bibinfo
  {author} {\bibfnamefont {J.~D.}\ \bibnamefont {Zimmerman}}, \ and\ \bibinfo
  {author} {\bibfnamefont {A.~C.}\ \bibnamefont {Gossard}},\ }\href@noop {}
  {\bibfield  {journal} {\bibinfo  {journal} {J. Cryst. Growth}\ }\textbf
  {\bibinfo {volume} {301}},\ \bibinfo {pages} {4} (\bibinfo {year}
  {2007})}\BibitemShut {NoStop}%
\bibitem [{\citenamefont {Bomberger}\ \emph {et~al.}(2016)\citenamefont
  {Bomberger}, \citenamefont {Tew}, \citenamefont {Lewis},\ and\ \citenamefont
  {Zide}}]{zide2016}%
  \BibitemOpen
  \bibfield  {author} {\bibinfo {author} {\bibfnamefont {C.~C.}\ \bibnamefont
  {Bomberger}}, \bibinfo {author} {\bibfnamefont {B.~E.}\ \bibnamefont {Tew}},
  \bibinfo {author} {\bibfnamefont {M.~R.}\ \bibnamefont {Lewis}}, \ and\
  \bibinfo {author} {\bibfnamefont {J.~M.~O.}\ \bibnamefont {Zide}},\
  }\href@noop {} {\bibfield  {journal} {\bibinfo  {journal} {Appl. Phys.
  Lett.}\ }\textbf {\bibinfo {volume} {109}},\ \bibinfo {pages} {202104}
  (\bibinfo {year} {2016})}\BibitemShut {NoStop}%
\bibitem [{\citenamefont {Bomberger}\ \emph {et~al.}(2017)\citenamefont
  {Bomberger}, \citenamefont {Lewis}, \citenamefont {Vanderhoef}, \citenamefont
  {Doty},\ and\ \citenamefont {Zide}}]{zide2017}%
  \BibitemOpen
  \bibfield  {author} {\bibinfo {author} {\bibfnamefont {C.~C.}\ \bibnamefont
  {Bomberger}}, \bibinfo {author} {\bibfnamefont {M.~R.}\ \bibnamefont
  {Lewis}}, \bibinfo {author} {\bibfnamefont {L.~R.}\ \bibnamefont
  {Vanderhoef}}, \bibinfo {author} {\bibfnamefont {M.~F.}\ \bibnamefont
  {Doty}}, \ and\ \bibinfo {author} {\bibfnamefont {J.~M.~O.}\ \bibnamefont
  {Zide}},\ }\href@noop {} {\bibfield  {journal} {\bibinfo  {journal} {â€ŽJ.
  Vac. Sci. Technol B}\ }\textbf {\bibinfo {volume} {35}},\ \bibinfo {pages}
  {030801} (\bibinfo {year} {2017})}\BibitemShut {NoStop}%
\bibitem [{\citenamefont {Zhang}\ \emph {et~al.}(2019)\citenamefont {Zhang},
  \citenamefont {Wang}, \citenamefont {Khalid}, \citenamefont {Janotti},
  \citenamefont {Haugstad},\ and\ \citenamefont {Zide}}]{Zhang2019}%
  \BibitemOpen
  \bibfield  {author} {\bibinfo {author} {\bibfnamefont {J.}~\bibnamefont
  {Zhang}}, \bibinfo {author} {\bibfnamefont {Y.}~\bibnamefont {Wang}},
  \bibinfo {author} {\bibfnamefont {S.}~\bibnamefont {Khalid}}, \bibinfo
  {author} {\bibfnamefont {A.}~\bibnamefont {Janotti}}, \bibinfo {author}
  {\bibfnamefont {G.}~\bibnamefont {Haugstad}}, \ and\ \bibinfo {author}
  {\bibfnamefont {J.~M.~O.}\ \bibnamefont {Zide}},\ }\href@noop {} {\bibfield
  {journal} {\bibinfo  {journal} {J. Appl. Phys}\ }\textbf {\bibinfo {volume}
  {126}},\ \bibinfo {pages} {095704} (\bibinfo {year} {2019})}\BibitemShut
  {NoStop}%
\bibitem [{\citenamefont {Niu}\ \emph {et~al.}(2016)\citenamefont {Niu},
  \citenamefont {Xu}, \citenamefont {Bai}, \citenamefont {Song}, \citenamefont
  {Shen}, \citenamefont {Xie}, \citenamefont {Sun}, \citenamefont {Huang},
  \citenamefont {Peets},\ and\ \citenamefont {Feng}}]{niu2016presence}%
  \BibitemOpen
  \bibfield  {author} {\bibinfo {author} {\bibfnamefont {X.~H.}\ \bibnamefont
  {Niu}}, \bibinfo {author} {\bibfnamefont {D.~F.}\ \bibnamefont {Xu}},
  \bibinfo {author} {\bibfnamefont {Y.~H.}\ \bibnamefont {Bai}}, \bibinfo
  {author} {\bibfnamefont {Q.}~\bibnamefont {Song}}, \bibinfo {author}
  {\bibfnamefont {X.~P.}\ \bibnamefont {Shen}}, \bibinfo {author}
  {\bibfnamefont {B.~P.}\ \bibnamefont {Xie}}, \bibinfo {author} {\bibfnamefont
  {Z.}~\bibnamefont {Sun}}, \bibinfo {author} {\bibfnamefont {Y.~B.}\
  \bibnamefont {Huang}}, \bibinfo {author} {\bibfnamefont {D.~C.}\ \bibnamefont
  {Peets}}, \ and\ \bibinfo {author} {\bibfnamefont {D.~L.}\ \bibnamefont
  {Feng}},\ }\href@noop {} {\bibfield  {journal} {\bibinfo  {journal} {Phys.
  Rev. B}\ }\textbf {\bibinfo {volume} {94}},\ \bibinfo {pages} {165163}
  (\bibinfo {year} {2016})}\BibitemShut {NoStop}%
\bibitem [{\citenamefont {Lou}\ \emph {et~al.}(2017)\citenamefont {Lou},
  \citenamefont {Fu}, \citenamefont {Xu}, \citenamefont {Guo}, \citenamefont
  {Kong}, \citenamefont {Zeng}, \citenamefont {Ma}, \citenamefont {Richard},
  \citenamefont {Fang}, \citenamefont {Huang}, \citenamefont {Sun},
  \citenamefont {Wang}, \citenamefont {Wang}, \citenamefont {Shi},
  \citenamefont {Lei}, \citenamefont {Liu}, \citenamefont {Weng}, \citenamefont
  {Qian}, \citenamefont {Ding},\ and\ \citenamefont {Wang}}]{lou2017evidence}%
  \BibitemOpen
  \bibfield  {author} {\bibinfo {author} {\bibfnamefont {R.}~\bibnamefont
  {Lou}}, \bibinfo {author} {\bibfnamefont {B.~B.}\ \bibnamefont {Fu}},
  \bibinfo {author} {\bibfnamefont {Q.~N.}\ \bibnamefont {Xu}}, \bibinfo
  {author} {\bibfnamefont {P.~J.}\ \bibnamefont {Guo}}, \bibinfo {author}
  {\bibfnamefont {L.~Y.}\ \bibnamefont {Kong}}, \bibinfo {author}
  {\bibfnamefont {L.~K.}\ \bibnamefont {Zeng}}, \bibinfo {author}
  {\bibfnamefont {J.~Z.}\ \bibnamefont {Ma}}, \bibinfo {author} {\bibfnamefont
  {P.}~\bibnamefont {Richard}}, \bibinfo {author} {\bibfnamefont
  {C.}~\bibnamefont {Fang}}, \bibinfo {author} {\bibfnamefont {Y.~B.}\
  \bibnamefont {Huang}}, \bibinfo {author} {\bibfnamefont {S.~S.}\ \bibnamefont
  {Sun}}, \bibinfo {author} {\bibfnamefont {Q.}~\bibnamefont {Wang}}, \bibinfo
  {author} {\bibfnamefont {L.}~\bibnamefont {Wang}}, \bibinfo {author}
  {\bibfnamefont {Y.~G.}\ \bibnamefont {Shi}}, \bibinfo {author} {\bibfnamefont
  {H.~C.}\ \bibnamefont {Lei}}, \bibinfo {author} {\bibfnamefont
  {K.}~\bibnamefont {Liu}}, \bibinfo {author} {\bibfnamefont {H.~M.}\
  \bibnamefont {Weng}}, \bibinfo {author} {\bibfnamefont {T.}~\bibnamefont
  {Qian}}, \bibinfo {author} {\bibfnamefont {H.}~\bibnamefont {Ding}}, \ and\
  \bibinfo {author} {\bibfnamefont {S.~C.}\ \bibnamefont {Wang}},\ }\href@noop
  {} {\bibfield  {journal} {\bibinfo  {journal} {Phys. Rev. B}\ }\textbf
  {\bibinfo {volume} {95}},\ \bibinfo {pages} {115140} (\bibinfo {year}
  {2017})}\BibitemShut {NoStop}%
\bibitem [{\citenamefont {Khalid}\ \emph {et~al.}(2018)\citenamefont {Khalid},
  \citenamefont {Sabino},\ and\ \citenamefont
  {Janotti}}]{khalid2018topological}%
  \BibitemOpen
  \bibfield  {author} {\bibinfo {author} {\bibfnamefont {S.}~\bibnamefont
  {Khalid}}, \bibinfo {author} {\bibfnamefont {F.~P.}\ \bibnamefont {Sabino}},
  \ and\ \bibinfo {author} {\bibfnamefont {A.}~\bibnamefont {Janotti}},\
  }\href@noop {} {\bibfield  {journal} {\bibinfo  {journal} {Phys. Rev. B}\
  }\textbf {\bibinfo {volume} {98}},\ \bibinfo {pages} {220102(R)} (\bibinfo
  {year} {2018})}\BibitemShut {NoStop}%
\bibitem [{\citenamefont {Zeng}\ \emph {et~al.}(2016)\citenamefont {Zeng},
  \citenamefont {Lou}, \citenamefont {Wu}, \citenamefont {Xu}, \citenamefont
  {Guo}, \citenamefont {Kong}, \citenamefont {Zhong}, \citenamefont {Ma},
  \citenamefont {Fu}, \citenamefont {Richard}, \citenamefont {Wang},
  \citenamefont {Liu}, \citenamefont {Lu}, \citenamefont {Huang}, \citenamefont
  {Fang}, \citenamefont {Sun}, \citenamefont {Wang}, \citenamefont {Wang},
  \citenamefont {Shi}, \citenamefont {Weng}, \citenamefont {Lei}, \citenamefont
  {Liu}, \citenamefont {Wang}, \citenamefont {Qian}, \citenamefont {Luo},\ and\
  \citenamefont {Ding}}]{zeng2016compensated}%
  \BibitemOpen
  \bibfield  {author} {\bibinfo {author} {\bibfnamefont {L.~K.}\ \bibnamefont
  {Zeng}}, \bibinfo {author} {\bibfnamefont {R.}~\bibnamefont {Lou}}, \bibinfo
  {author} {\bibfnamefont {D.~S.}\ \bibnamefont {Wu}}, \bibinfo {author}
  {\bibfnamefont {Q.~N.}\ \bibnamefont {Xu}}, \bibinfo {author} {\bibfnamefont
  {P.~J.}\ \bibnamefont {Guo}}, \bibinfo {author} {\bibfnamefont {L.~Y.}\
  \bibnamefont {Kong}}, \bibinfo {author} {\bibfnamefont {Y.~G.}\ \bibnamefont
  {Zhong}}, \bibinfo {author} {\bibfnamefont {J.~Z.}\ \bibnamefont {Ma}},
  \bibinfo {author} {\bibfnamefont {B.~B.}\ \bibnamefont {Fu}}, \bibinfo
  {author} {\bibfnamefont {P.}~\bibnamefont {Richard}}, \bibinfo {author}
  {\bibfnamefont {P.}~\bibnamefont {Wang}}, \bibinfo {author} {\bibfnamefont
  {G.~T.}\ \bibnamefont {Liu}}, \bibinfo {author} {\bibfnamefont
  {L.}~\bibnamefont {Lu}}, \bibinfo {author} {\bibfnamefont {Y.~B.}\
  \bibnamefont {Huang}}, \bibinfo {author} {\bibfnamefont {C.}~\bibnamefont
  {Fang}}, \bibinfo {author} {\bibfnamefont {S.~S.}\ \bibnamefont {Sun}},
  \bibinfo {author} {\bibfnamefont {Q.}~\bibnamefont {Wang}}, \bibinfo {author}
  {\bibfnamefont {L.}~\bibnamefont {Wang}}, \bibinfo {author} {\bibfnamefont
  {Y.~G.}\ \bibnamefont {Shi}}, \bibinfo {author} {\bibfnamefont {H.~M.}\
  \bibnamefont {Weng}}, \bibinfo {author} {\bibfnamefont {H.~C.}\ \bibnamefont
  {Lei}}, \bibinfo {author} {\bibfnamefont {K.}~\bibnamefont {Liu}}, \bibinfo
  {author} {\bibfnamefont {S.~C.}\ \bibnamefont {Wang}}, \bibinfo {author}
  {\bibfnamefont {T.}~\bibnamefont {Qian}}, \bibinfo {author} {\bibfnamefont
  {J.~L.}\ \bibnamefont {Luo}}, \ and\ \bibinfo {author} {\bibfnamefont
  {H.}~\bibnamefont {Ding}},\ }\href@noop {} {\bibfield  {journal} {\bibinfo
  {journal} {Phys. Rev. Lett.}\ }\textbf {\bibinfo {volume} {117}},\ \bibinfo
  {pages} {127204} (\bibinfo {year} {2016})}\BibitemShut {NoStop}%
\bibitem [{\citenamefont {Guo}\ \emph {et~al.}(2016)\citenamefont {Guo},
  \citenamefont {Yang}, \citenamefont {Zhang}, \citenamefont {Liu},\ and\
  \citenamefont {Lu}}]{guo2016charge}%
  \BibitemOpen
  \bibfield  {author} {\bibinfo {author} {\bibfnamefont {P.~J.}\ \bibnamefont
  {Guo}}, \bibinfo {author} {\bibfnamefont {H.-C.}\ \bibnamefont {Yang}},
  \bibinfo {author} {\bibfnamefont {B.~J.}\ \bibnamefont {Zhang}}, \bibinfo
  {author} {\bibfnamefont {K.}~\bibnamefont {Liu}}, \ and\ \bibinfo {author}
  {\bibfnamefont {Z.~Y.}\ \bibnamefont {Lu}},\ }\href@noop {} {\bibfield
  {journal} {\bibinfo  {journal} {Phys. Rev. B}\ }\textbf {\bibinfo {volume}
  {93}},\ \bibinfo {pages} {235142} (\bibinfo {year} {2016})}\BibitemShut
  {NoStop}%
\bibitem [{\citenamefont {Tafti}\ \emph {et~al.}(2017)\citenamefont {Tafti},
  \citenamefont {Torikachvili}, \citenamefont {Stillwell}, \citenamefont
  {Baer}, \citenamefont {Stavrou}, \citenamefont {Weir}, \citenamefont {Vohra},
  \citenamefont {Yang}, \citenamefont {McDonnell}, \citenamefont {Kushwaha},
  \citenamefont {Gibson}, \citenamefont {Cava},\ and\ \citenamefont
  {Jeffries}}]{tafti2017tuning}%
  \BibitemOpen
  \bibfield  {author} {\bibinfo {author} {\bibfnamefont {F.~F.}\ \bibnamefont
  {Tafti}}, \bibinfo {author} {\bibfnamefont {M.~S.}\ \bibnamefont
  {Torikachvili}}, \bibinfo {author} {\bibfnamefont {R.~L.}\ \bibnamefont
  {Stillwell}}, \bibinfo {author} {\bibfnamefont {B.}~\bibnamefont {Baer}},
  \bibinfo {author} {\bibfnamefont {E.}~\bibnamefont {Stavrou}}, \bibinfo
  {author} {\bibfnamefont {S.~T.}\ \bibnamefont {Weir}}, \bibinfo {author}
  {\bibfnamefont {Y.~K.}\ \bibnamefont {Vohra}}, \bibinfo {author}
  {\bibfnamefont {H.-Y.}\ \bibnamefont {Yang}}, \bibinfo {author}
  {\bibfnamefont {E.~F.}\ \bibnamefont {McDonnell}}, \bibinfo {author}
  {\bibfnamefont {S.~K.}\ \bibnamefont {Kushwaha}}, \bibinfo {author}
  {\bibfnamefont {Q.~D.}\ \bibnamefont {Gibson}}, \bibinfo {author}
  {\bibfnamefont {R.~J.}\ \bibnamefont {Cava}}, \ and\ \bibinfo {author}
  {\bibfnamefont {J.~R.}\ \bibnamefont {Jeffries}},\ }\href@noop {} {\bibfield
  {journal} {\bibinfo  {journal} {Phys. Rev. B}\ }\textbf {\bibinfo {volume}
  {95}},\ \bibinfo {pages} {014507} (\bibinfo {year} {2017})}\BibitemShut
  {NoStop}%
\bibitem [{\citenamefont {Allen~Jr}\ \emph {et~al.}(1989)\citenamefont
  {Allen~Jr}, \citenamefont {Tabatabaie}, \citenamefont {Palmstr\o{}m},
  \citenamefont {Hull}, \citenamefont {Sands}, \citenamefont {DeRosa},
  \citenamefont {Gilchrist},\ and\ \citenamefont {Garrison}}]{allen1989eras}%
  \BibitemOpen
  \bibfield  {author} {\bibinfo {author} {\bibfnamefont {S.~J.}\ \bibnamefont
  {Allen~Jr}}, \bibinfo {author} {\bibfnamefont {N.}~\bibnamefont
  {Tabatabaie}}, \bibinfo {author} {\bibfnamefont {C.~J.}\ \bibnamefont
  {Palmstr\o{}m}}, \bibinfo {author} {\bibfnamefont {G.~W.}\ \bibnamefont
  {Hull}}, \bibinfo {author} {\bibfnamefont {T.}~\bibnamefont {Sands}},
  \bibinfo {author} {\bibfnamefont {F.}~\bibnamefont {DeRosa}}, \bibinfo
  {author} {\bibfnamefont {H.~L.}\ \bibnamefont {Gilchrist}}, \ and\ \bibinfo
  {author} {\bibfnamefont {K.~C.}\ \bibnamefont {Garrison}},\ }\href@noop {}
  {\bibfield  {journal} {\bibinfo  {journal} {Phys. Rev. Lett.}\ }\textbf
  {\bibinfo {volume} {62}},\ \bibinfo {pages} {2309} (\bibinfo {year}
  {1989})}\BibitemShut {NoStop}%
\bibitem [{\citenamefont {Abdusalyamova}\ and\ \citenamefont
  {Rachmatov}(2002)}]{abdusalyamova2002investigation}%
  \BibitemOpen
  \bibfield  {author} {\bibinfo {author} {\bibfnamefont {M.~N.}\ \bibnamefont
  {Abdusalyamova}}\ and\ \bibinfo {author} {\bibfnamefont {O.~I.}\ \bibnamefont
  {Rachmatov}},\ }\href@noop {} {\bibfield  {journal} {\bibinfo  {journal} {Z.
  Naturforsch. A}\ }\textbf {\bibinfo {volume} {57}},\ \bibinfo {pages} {98}
  (\bibinfo {year} {2002})}\BibitemShut {NoStop}%
\bibitem [{\citenamefont {Child}\ \emph {et~al.}(1963)\citenamefont {Child},
  \citenamefont {Wilkinson}, \citenamefont {Cable}, \citenamefont {Koehler},\
  and\ \citenamefont {Wollan}}]{child1963neutron}%
  \BibitemOpen
  \bibfield  {author} {\bibinfo {author} {\bibfnamefont {H.~R.}\ \bibnamefont
  {Child}}, \bibinfo {author} {\bibfnamefont {M.~K.}\ \bibnamefont
  {Wilkinson}}, \bibinfo {author} {\bibfnamefont {J.~W.}\ \bibnamefont
  {Cable}}, \bibinfo {author} {\bibfnamefont {W.~C.}\ \bibnamefont {Koehler}},
  \ and\ \bibinfo {author} {\bibfnamefont {E.~O.}\ \bibnamefont {Wollan}},\
  }\href@noop {} {\bibfield  {journal} {\bibinfo  {journal} {Phys. Rev.}\
  }\textbf {\bibinfo {volume} {131}},\ \bibinfo {pages} {922} (\bibinfo {year}
  {1963})}\BibitemShut {NoStop}%
\bibitem [{\citenamefont {Petit}\ \emph {et~al.}(2016)\citenamefont {Petit},
  \citenamefont {Szotek}, \citenamefont {L{\"u}ders},\ and\ \citenamefont
  {Svane}}]{petit2016rare}%
  \BibitemOpen
  \bibfield  {author} {\bibinfo {author} {\bibfnamefont {L.}~\bibnamefont
  {Petit}}, \bibinfo {author} {\bibfnamefont {Z.}~\bibnamefont {Szotek}},
  \bibinfo {author} {\bibfnamefont {M.}~\bibnamefont {L{\"u}ders}}, \ and\
  \bibinfo {author} {\bibfnamefont {A.}~\bibnamefont {Svane}},\ }\href@noop {}
  {\bibfield  {journal} {\bibinfo  {journal} {J. Phys. Condens. Matter}\
  }\textbf {\bibinfo {volume} {28}},\ \bibinfo {pages} {223001} (\bibinfo
  {year} {2016})}\BibitemShut {NoStop}%
\bibitem [{\citenamefont {Sclar}(1962)}]{sclar1962energy}%
  \BibitemOpen
  \bibfield  {author} {\bibinfo {author} {\bibfnamefont {N.}~\bibnamefont
  {Sclar}},\ }\href@noop {} {\bibfield  {journal} {\bibinfo  {journal} {J.
  Appl. Phys.}\ }\textbf {\bibinfo {volume} {33}},\ \bibinfo {pages} {2999}
  (\bibinfo {year} {1962})}\BibitemShut {NoStop}%
\bibitem [{\citenamefont {Andersen}(1980)}]{Andersen1980}%
  \BibitemOpen
  \bibfield  {author} {\bibinfo {author} {\bibfnamefont {N.~H.}\ \bibnamefont
  {Andersen}},\ }\href@noop {} {\emph {\bibinfo {title} {Crystalline electric
  field and structural effects in f-electron systems}}}\ (\bibinfo  {publisher}
  {Springer},\ \bibinfo {year} {1980})\ pp.\ \bibinfo {pages}
  {373--387}\BibitemShut {NoStop}%
\bibitem [{\citenamefont {Eyring}(1999)}]{Eyring1999}%
  \BibitemOpen
  \bibfield  {author} {\bibinfo {author} {\bibfnamefont {L.}~\bibnamefont
  {Eyring}},\ }\href@noop {} {\emph {\bibinfo {title} {Handbook on the Physics
  and Chemistry of Rare Earths}}},\ Vol.~\bibinfo {volume} {26}\ (\bibinfo
  {publisher} {Elsevier},\ \bibinfo {year} {1999})\BibitemShut {NoStop}%
\bibitem [{\citenamefont {Trammell}(1963)}]{trammell1963magnetic}%
  \BibitemOpen
  \bibfield  {author} {\bibinfo {author} {\bibfnamefont {G.~T.}\ \bibnamefont
  {Trammell}},\ }\href@noop {} {\bibfield  {journal} {\bibinfo  {journal}
  {Phys. Rev.}\ }\textbf {\bibinfo {volume} {131}},\ \bibinfo {pages} {932}
  (\bibinfo {year} {1963})}\BibitemShut {NoStop}%
\bibitem [{\citenamefont {H.Takahashi}\ and\ \citenamefont
  {T.Kasuya}(1985)}]{Takahashi1985}%
  \BibitemOpen
  \bibfield  {author} {\bibinfo {author} {\bibnamefont {H.Takahashi}}\ and\
  \bibinfo {author} {\bibnamefont {T.Kasuya}},\ }\href
  {http://stacks.iop.org/0022-3719/18/i=13/a=016} {\bibfield  {journal}
  {\bibinfo  {journal} {J. Phys. C. Solid. State. Phys}\ }\textbf {\bibinfo
  {volume} {18}},\ \bibinfo {pages} {2697} (\bibinfo {year}
  {1985})}\BibitemShut {NoStop}%
\bibitem [{\citenamefont {Narita}\ and\ \citenamefont
  {Kasuya}(1985)}]{narita1985magnetic}%
  \BibitemOpen
  \bibfield  {author} {\bibinfo {author} {\bibfnamefont {A.}~\bibnamefont
  {Narita}}\ and\ \bibinfo {author} {\bibfnamefont {T.}~\bibnamefont
  {Kasuya}},\ }\href@noop {} {\bibfield  {journal} {\bibinfo  {journal} {J.
  Magn. Magn. Mater.}\ }\textbf {\bibinfo {volume} {52}},\ \bibinfo {pages}
  {373} (\bibinfo {year} {1985})}\BibitemShut {NoStop}%
\bibitem [{\citenamefont {Birgeneau}\ \emph {et~al.}(1973)\citenamefont
  {Birgeneau}, \citenamefont {Bucher}, \citenamefont {Maita}, \citenamefont
  {Passell},\ and\ \citenamefont {Turberfield}}]{birgeneau1973crystal}%
  \BibitemOpen
  \bibfield  {author} {\bibinfo {author} {\bibfnamefont {R.~J.}\ \bibnamefont
  {Birgeneau}}, \bibinfo {author} {\bibfnamefont {E.}~\bibnamefont {Bucher}},
  \bibinfo {author} {\bibfnamefont {J.~P.}\ \bibnamefont {Maita}}, \bibinfo
  {author} {\bibfnamefont {L.}~\bibnamefont {Passell}}, \ and\ \bibinfo
  {author} {\bibfnamefont {K.~C.}\ \bibnamefont {Turberfield}},\ }\href@noop {}
  {\bibfield  {journal} {\bibinfo  {journal} {Phys. Rev. B}\ }\textbf {\bibinfo
  {volume} {8}},\ \bibinfo {pages} {5345} (\bibinfo {year} {1973})}\BibitemShut
  {NoStop}%
\bibitem [{\citenamefont {Hasegawa}\ and\ \citenamefont
  {Yanase}(1977)}]{hasegawa1977energy}%
  \BibitemOpen
  \bibfield  {author} {\bibinfo {author} {\bibfnamefont {A.}~\bibnamefont
  {Hasegawa}}\ and\ \bibinfo {author} {\bibfnamefont {A.}~\bibnamefont
  {Yanase}},\ }\href@noop {} {\bibfield  {journal} {\bibinfo  {journal} {J.
  Phys. Soc. Jpn.}\ }\textbf {\bibinfo {volume} {42}},\ \bibinfo {pages} {492}
  (\bibinfo {year} {1977})}\BibitemShut {NoStop}%
\bibitem [{\citenamefont {Slater}(1951)}]{slater1951simplification}%
  \BibitemOpen
  \bibfield  {author} {\bibinfo {author} {\bibfnamefont {J.~C.}\ \bibnamefont
  {Slater}},\ }\href@noop {} {\bibfield  {journal} {\bibinfo  {journal} {Phys.
  Rev.}\ }\textbf {\bibinfo {volume} {81}},\ \bibinfo {pages} {385} (\bibinfo
  {year} {1951})}\BibitemShut {NoStop}%
\bibitem [{\citenamefont {Petukhov}\ \emph {et~al.}(1994)\citenamefont
  {Petukhov}, \citenamefont {Lambrecht},\ and\ \citenamefont
  {Segall}}]{petukhov1994electronic}%
  \BibitemOpen
  \bibfield  {author} {\bibinfo {author} {\bibfnamefont {A.~G.}\ \bibnamefont
  {Petukhov}}, \bibinfo {author} {\bibfnamefont {W.~R.~L.}\ \bibnamefont
  {Lambrecht}}, \ and\ \bibinfo {author} {\bibfnamefont {B.}~\bibnamefont
  {Segall}},\ }\href@noop {} {\bibfield  {journal} {\bibinfo  {journal} {Phys.
  Rev. B}\ }\textbf {\bibinfo {volume} {50}},\ \bibinfo {pages} {7800}
  (\bibinfo {year} {1994})}\BibitemShut {NoStop}%
\bibitem [{\citenamefont {Petukhov}\ \emph {et~al.}(1996)\citenamefont
  {Petukhov}, \citenamefont {Lambrecht},\ and\ \citenamefont
  {Segall}}]{petukhov1996electronic}%
  \BibitemOpen
  \bibfield  {author} {\bibinfo {author} {\bibfnamefont {A.~G.}\ \bibnamefont
  {Petukhov}}, \bibinfo {author} {\bibfnamefont {W.~R.~L.}\ \bibnamefont
  {Lambrecht}}, \ and\ \bibinfo {author} {\bibfnamefont {B.}~\bibnamefont
  {Segall}},\ }\href@noop {} {\bibfield  {journal} {\bibinfo  {journal} {Phys.
  Rev. B}\ }\textbf {\bibinfo {volume} {53}},\ \bibinfo {pages} {4324}
  (\bibinfo {year} {1996})}\BibitemShut {NoStop}%
\bibitem [{\citenamefont {Brooks}\ \emph {et~al.}(1991)\citenamefont {Brooks},
  \citenamefont {Nordstrom},\ and\ \citenamefont {Johansson}}]{brooks19913d}%
  \BibitemOpen
  \bibfield  {author} {\bibinfo {author} {\bibfnamefont {M.~S.~S.}\
  \bibnamefont {Brooks}}, \bibinfo {author} {\bibfnamefont {L.}~\bibnamefont
  {Nordstrom}}, \ and\ \bibinfo {author} {\bibfnamefont {B.}~\bibnamefont
  {Johansson}},\ }\href@noop {} {\bibfield  {journal} {\bibinfo  {journal} {J.
  Phys. Condens. Matter}\ }\textbf {\bibinfo {volume} {3}},\ \bibinfo {pages}
  {2357} (\bibinfo {year} {1991})}\BibitemShut {NoStop}%
\bibitem [{\citenamefont {Temmerman}\ and\ \citenamefont
  {Sterne}(1990)}]{temmerman1990band}%
  \BibitemOpen
  \bibfield  {author} {\bibinfo {author} {\bibfnamefont {W.~M.}\ \bibnamefont
  {Temmerman}}\ and\ \bibinfo {author} {\bibfnamefont {P.~A.}\ \bibnamefont
  {Sterne}},\ }\href@noop {} {\bibfield  {journal} {\bibinfo  {journal} {J.
  Phys. Condens. Matter}\ }\textbf {\bibinfo {volume} {2}},\ \bibinfo {pages}
  {5529} (\bibinfo {year} {1990})}\BibitemShut {NoStop}%
\bibitem [{\citenamefont {Heinemann}\ and\ \citenamefont
  {Temmerman}(1994)}]{heinemann1994magnetic}%
  \BibitemOpen
  \bibfield  {author} {\bibinfo {author} {\bibfnamefont {M.}~\bibnamefont
  {Heinemann}}\ and\ \bibinfo {author} {\bibfnamefont {W.~M.}\ \bibnamefont
  {Temmerman}},\ }\href@noop {} {\bibfield  {journal} {\bibinfo  {journal}
  {Phys. Rev. B}\ }\textbf {\bibinfo {volume} {49}},\ \bibinfo {pages} {4348}
  (\bibinfo {year} {1994})}\BibitemShut {NoStop}%
\bibitem [{\citenamefont {Pourovskii}\ \emph {et~al.}(2009)\citenamefont
  {Pourovskii}, \citenamefont {Delaney}, \citenamefont {Van~de Walle},
  \citenamefont {Spaldin},\ and\ \citenamefont {Georges}}]{pourovskii2009role}%
  \BibitemOpen
  \bibfield  {author} {\bibinfo {author} {\bibfnamefont {L.~V.}\ \bibnamefont
  {Pourovskii}}, \bibinfo {author} {\bibfnamefont {K.~T.}\ \bibnamefont
  {Delaney}}, \bibinfo {author} {\bibfnamefont {C.~G.}\ \bibnamefont {Van~de
  Walle}}, \bibinfo {author} {\bibfnamefont {N.~A.}\ \bibnamefont {Spaldin}}, \
  and\ \bibinfo {author} {\bibfnamefont {A.}~\bibnamefont {Georges}},\
  }\href@noop {} {\bibfield  {journal} {\bibinfo  {journal} {Phys. Rev. Lett.}\
  }\textbf {\bibinfo {volume} {102}},\ \bibinfo {pages} {096401} (\bibinfo
  {year} {2009})}\BibitemShut {NoStop}%
\bibitem [{\citenamefont {Hohenberg}\ and\ \citenamefont
  {Kohn}(1964)}]{hohenberg1964inhomogeneous}%
  \BibitemOpen
  \bibfield  {author} {\bibinfo {author} {\bibfnamefont {P.}~\bibnamefont
  {Hohenberg}}\ and\ \bibinfo {author} {\bibfnamefont {W.}~\bibnamefont
  {Kohn}},\ }\href@noop {} {\bibfield  {journal} {\bibinfo  {journal} {Phys.
  Rev}\ }\textbf {\bibinfo {volume} {136}},\ \bibinfo {pages} {B864} (\bibinfo
  {year} {1964})}\BibitemShut {NoStop}%
\bibitem [{\citenamefont {Kohn}\ and\ \citenamefont
  {Sham}(1965)}]{kohn1965self}%
  \BibitemOpen
  \bibfield  {author} {\bibinfo {author} {\bibfnamefont {W.}~\bibnamefont
  {Kohn}}\ and\ \bibinfo {author} {\bibfnamefont {L.~J.}\ \bibnamefont
  {Sham}},\ }\href@noop {} {\bibfield  {journal} {\bibinfo  {journal} {Phys.
  Rev}\ }\textbf {\bibinfo {volume} {140}},\ \bibinfo {pages} {A1133} (\bibinfo
  {year} {1965})}\BibitemShut {NoStop}%
\bibitem [{\citenamefont {Heyd}\ \emph {et~al.}(2003)\citenamefont {Heyd},
  \citenamefont {Scuseria},\ and\ \citenamefont {Ernzerhof}}]{heyd2003hybrid}%
  \BibitemOpen
  \bibfield  {author} {\bibinfo {author} {\bibfnamefont {J.}~\bibnamefont
  {Heyd}}, \bibinfo {author} {\bibfnamefont {G.~E.}\ \bibnamefont {Scuseria}},
  \ and\ \bibinfo {author} {\bibfnamefont {M.}~\bibnamefont {Ernzerhof}},\
  }\href@noop {} {\bibfield  {journal} {\bibinfo  {journal} {J. Chem. Phys}\
  }\textbf {\bibinfo {volume} {118}},\ \bibinfo {pages} {8207} (\bibinfo {year}
  {2003})}\BibitemShut {NoStop}%
\bibitem [{\citenamefont {Heyd}\ \emph {et~al.}(2006)\citenamefont {Heyd},
  \citenamefont {Scuseria},\ and\ \citenamefont {Ernzerhof}}]{HSE}%
  \BibitemOpen
  \bibfield  {author} {\bibinfo {author} {\bibfnamefont {J.}~\bibnamefont
  {Heyd}}, \bibinfo {author} {\bibfnamefont {G.~E.}\ \bibnamefont {Scuseria}},
  \ and\ \bibinfo {author} {\bibfnamefont {M.}~\bibnamefont {Ernzerhof}},\
  }\href {\doibase 10.1063/1.2204597} {\bibfield  {journal} {\bibinfo
  {journal} {J. Chem. Phys.}\ }\textbf {\bibinfo {volume} {124}},\ \bibinfo
  {pages} {219906} (\bibinfo {year} {2006})}\BibitemShut {NoStop}%
\bibitem [{\citenamefont {Kresse}\ and\ \citenamefont
  {Hafner}(1993)}]{kresse1993ab}%
  \BibitemOpen
  \bibfield  {author} {\bibinfo {author} {\bibfnamefont {G.}~\bibnamefont
  {Kresse}}\ and\ \bibinfo {author} {\bibfnamefont {J.}~\bibnamefont
  {Hafner}},\ }\href@noop {} {\bibfield  {journal} {\bibinfo  {journal} {Phys.
  Rev. B}\ }\textbf {\bibinfo {volume} {47}},\ \bibinfo {pages} {558} (\bibinfo
  {year} {1993})}\BibitemShut {NoStop}%
\bibitem [{\citenamefont {Kresse}\ and\ \citenamefont
  {Hafner}(1994)}]{kresse1994ab}%
  \BibitemOpen
  \bibfield  {author} {\bibinfo {author} {\bibfnamefont {G.}~\bibnamefont
  {Kresse}}\ and\ \bibinfo {author} {\bibfnamefont {J.}~\bibnamefont
  {Hafner}},\ }\href@noop {} {\bibfield  {journal} {\bibinfo  {journal} {Phys.
  Rev. B}\ }\textbf {\bibinfo {volume} {49}},\ \bibinfo {pages} {14251}
  (\bibinfo {year} {1994})}\BibitemShut {NoStop}%
\bibitem [{\citenamefont {Perdew}\ \emph {et~al.}(1996)\citenamefont {Perdew},
  \citenamefont {Burke},\ and\ \citenamefont
  {Ernzerhof}}]{perdew1996generalized}%
  \BibitemOpen
  \bibfield  {author} {\bibinfo {author} {\bibfnamefont {J.~P.}\ \bibnamefont
  {Perdew}}, \bibinfo {author} {\bibfnamefont {K.}~\bibnamefont {Burke}}, \
  and\ \bibinfo {author} {\bibfnamefont {M.}~\bibnamefont {Ernzerhof}},\
  }\href@noop {} {\bibfield  {journal} {\bibinfo  {journal} {Phys. Rev. Lett.}\
  }\textbf {\bibinfo {volume} {77}},\ \bibinfo {pages} {3865} (\bibinfo {year}
  {1996})}\BibitemShut {NoStop}%
\bibitem [{\citenamefont {Bl{\"o}chl}(1994)}]{blochl1994projector}%
  \BibitemOpen
  \bibfield  {author} {\bibinfo {author} {\bibfnamefont {P.~E.}\ \bibnamefont
  {Bl{\"o}chl}},\ }\href@noop {} {\bibfield  {journal} {\bibinfo  {journal}
  {Phys. Rev. B}\ }\textbf {\bibinfo {volume} {50}},\ \bibinfo {pages} {17953}
  (\bibinfo {year} {1994})}\BibitemShut {NoStop}%
\bibitem [{\citenamefont {Li}\ \emph {et~al.}(1997)\citenamefont {Li},
  \citenamefont {Haga}, \citenamefont {Shida}, \citenamefont {Suzuki},
  \citenamefont {Kwon},\ and\ \citenamefont {Kido}}]{li1997magnetic}%
  \BibitemOpen
  \bibfield  {author} {\bibinfo {author} {\bibfnamefont {D.~X.}\ \bibnamefont
  {Li}}, \bibinfo {author} {\bibfnamefont {Y.}~\bibnamefont {Haga}}, \bibinfo
  {author} {\bibfnamefont {H.}~\bibnamefont {Shida}}, \bibinfo {author}
  {\bibfnamefont {T.}~\bibnamefont {Suzuki}}, \bibinfo {author} {\bibfnamefont
  {Y.~S.}\ \bibnamefont {Kwon}}, \ and\ \bibinfo {author} {\bibfnamefont
  {G.}~\bibnamefont {Kido}},\ }\href@noop {} {\bibfield  {journal} {\bibinfo
  {journal} {J. Phys. Condens. Matter}\ }\textbf {\bibinfo {volume} {9}},\
  \bibinfo {pages} {10777} (\bibinfo {year} {1997})}\BibitemShut {NoStop}%
\bibitem [{\citenamefont {Allen~Jr}\ \emph {et~al.}(1990)\citenamefont
  {Allen~Jr}, \citenamefont {Tabatabaie}, \citenamefont {Palmstr\o{}m},
  \citenamefont {Mounier}, \citenamefont {Hull}, \citenamefont {Sands},
  \citenamefont {DeRosa}, \citenamefont {Gilchrist},\ and\ \citenamefont
  {Garrison}}]{allen1990magneto}%
  \BibitemOpen
  \bibfield  {author} {\bibinfo {author} {\bibfnamefont {S.}~\bibnamefont
  {Allen~Jr}}, \bibinfo {author} {\bibfnamefont {N.}~\bibnamefont
  {Tabatabaie}}, \bibinfo {author} {\bibfnamefont {C.~J.}\ \bibnamefont
  {Palmstr\o{}m}}, \bibinfo {author} {\bibfnamefont {S.}~\bibnamefont
  {Mounier}}, \bibinfo {author} {\bibfnamefont {G.~W.}\ \bibnamefont {Hull}},
  \bibinfo {author} {\bibfnamefont {T.}~\bibnamefont {Sands}}, \bibinfo
  {author} {\bibfnamefont {F.}~\bibnamefont {DeRosa}}, \bibinfo {author}
  {\bibfnamefont {H.~L.}\ \bibnamefont {Gilchrist}}, \ and\ \bibinfo {author}
  {\bibfnamefont {K.~C.}\ \bibnamefont {Garrison}},\ }\href@noop {} {\bibfield
  {journal} {\bibinfo  {journal} {Surf. Sci.}\ }\textbf {\bibinfo {volume}
  {228}},\ \bibinfo {pages} {13} (\bibinfo {year} {1990})}\BibitemShut
  {NoStop}%
\bibitem [{\citenamefont {Mullen}\ \emph {et~al.}(1974)\citenamefont {Mullen},
  \citenamefont {L{\"u}thi}, \citenamefont {Wang}, \citenamefont {Bucher},
  \citenamefont {Longinotti}, \citenamefont {Maita},\ and\ \citenamefont
  {Ott}}]{mullen1974magnetic}%
  \BibitemOpen
  \bibfield  {author} {\bibinfo {author} {\bibfnamefont {M.~E.}\ \bibnamefont
  {Mullen}}, \bibinfo {author} {\bibfnamefont {B.}~\bibnamefont {L{\"u}thi}},
  \bibinfo {author} {\bibfnamefont {P.~S.}\ \bibnamefont {Wang}}, \bibinfo
  {author} {\bibfnamefont {E.}~\bibnamefont {Bucher}}, \bibinfo {author}
  {\bibfnamefont {L.~D.}\ \bibnamefont {Longinotti}}, \bibinfo {author}
  {\bibfnamefont {J.~P.}\ \bibnamefont {Maita}}, \ and\ \bibinfo {author}
  {\bibfnamefont {H.~R.}\ \bibnamefont {Ott}},\ }\href@noop {} {\bibfield
  {journal} {\bibinfo  {journal} {Phys. Rev. B}\ }\textbf {\bibinfo {volume}
  {10}},\ \bibinfo {pages} {186} (\bibinfo {year} {1974})}\BibitemShut
  {NoStop}%
\bibitem [{\citenamefont {Chua}\ and\ \citenamefont
  {Pratt}(1974)}]{chua1974simple}%
  \BibitemOpen
  \bibfield  {author} {\bibinfo {author} {\bibfnamefont {K.~S.}\ \bibnamefont
  {Chua}}\ and\ \bibinfo {author} {\bibfnamefont {J.~N.}\ \bibnamefont
  {Pratt}},\ }\href@noop {} {\bibfield  {journal} {\bibinfo  {journal}
  {Thermochim Acta}\ }\textbf {\bibinfo {volume} {8}},\ \bibinfo {pages} {409}
  (\bibinfo {year} {1974})}\BibitemShut {NoStop}%
\bibitem [{\citenamefont {Shirotani}\ \emph {et~al.}(2003)\citenamefont
  {Shirotani}, \citenamefont {Yamanashi}, \citenamefont {Hayashi},
  \citenamefont {Ishimatsu}, \citenamefont {Shimomura},\ and\ \citenamefont
  {Kikegawa}}]{shirotani2003pressure}%
  \BibitemOpen
  \bibfield  {author} {\bibinfo {author} {\bibfnamefont {I.}~\bibnamefont
  {Shirotani}}, \bibinfo {author} {\bibfnamefont {K.}~\bibnamefont
  {Yamanashi}}, \bibinfo {author} {\bibfnamefont {J.}~\bibnamefont {Hayashi}},
  \bibinfo {author} {\bibfnamefont {N.}~\bibnamefont {Ishimatsu}}, \bibinfo
  {author} {\bibfnamefont {O.}~\bibnamefont {Shimomura}}, \ and\ \bibinfo
  {author} {\bibfnamefont {T.}~\bibnamefont {Kikegawa}},\ }\href@noop {}
  {\bibfield  {journal} {\bibinfo  {journal} {Solid State Commun.}\ }\textbf
  {\bibinfo {volume} {127}},\ \bibinfo {pages} {573} (\bibinfo {year}
  {2003})}\BibitemShut {NoStop}%
\bibitem [{\citenamefont {Cotton}\ \emph {et~al.}(1988)\citenamefont {Cotton},
  \citenamefont {Wilkinson}, \citenamefont {Murillo},\ and\ \citenamefont
  {Bochmann}}]{cotton1988advanced}%
  \BibitemOpen
  \bibfield  {author} {\bibinfo {author} {\bibfnamefont {F.~A.}\ \bibnamefont
  {Cotton}}, \bibinfo {author} {\bibfnamefont {G.}~\bibnamefont {Wilkinson}},
  \bibinfo {author} {\bibfnamefont {C.~A.}\ \bibnamefont {Murillo}}, \ and\
  \bibinfo {author} {\bibfnamefont {M.}~\bibnamefont {Bochmann}},\ }\href@noop
  {} {\emph {\bibinfo {title} {{Advanced inorganic chemistry}}}},\
  Vol.~\bibinfo {volume} {6}\ (\bibinfo  {publisher} {Wiley New York},\
  \bibinfo {year} {1988})\BibitemShut {NoStop}%
\bibitem [{\citenamefont {Harrison}(2012)}]{harrison2012electronic}%
  \BibitemOpen
  \bibfield  {author} {\bibinfo {author} {\bibfnamefont {W.~A.}\ \bibnamefont
  {Harrison}},\ }\href@noop {} {\emph {\bibinfo {title} {{Electronic structure
  and the properties of solids: the physics of the chemical bond}}}}\ (\bibinfo
   {publisher} {Courier Corporation},\ \bibinfo {year} {2012})\BibitemShut
  {NoStop}%
\bibitem [{\citenamefont {Duan}\ \emph {et~al.}(2004)\citenamefont {Duan},
  \citenamefont {Komesu}, \citenamefont {Jeong}, \citenamefont {Borca},
  \citenamefont {Yin}, \citenamefont {Liu}, \citenamefont {Mei}, \citenamefont
  {Dowben}, \citenamefont {Petukhov}, \citenamefont {Schultz},\ and\
  \citenamefont {Palmstr\o{}m}}]{duan2004hybridization}%
  \BibitemOpen
  \bibfield  {author} {\bibinfo {author} {\bibfnamefont {C.~G.}\ \bibnamefont
  {Duan}}, \bibinfo {author} {\bibfnamefont {T.}~\bibnamefont {Komesu}},
  \bibinfo {author} {\bibfnamefont {H.~K.}\ \bibnamefont {Jeong}}, \bibinfo
  {author} {\bibfnamefont {C.~N.}\ \bibnamefont {Borca}}, \bibinfo {author}
  {\bibfnamefont {W.~G.}\ \bibnamefont {Yin}}, \bibinfo {author} {\bibfnamefont
  {J.}~\bibnamefont {Liu}}, \bibinfo {author} {\bibfnamefont {W.~N.}\
  \bibnamefont {Mei}}, \bibinfo {author} {\bibfnamefont {P.~A.}\ \bibnamefont
  {Dowben}}, \bibinfo {author} {\bibfnamefont {A.~G.}\ \bibnamefont
  {Petukhov}}, \bibinfo {author} {\bibfnamefont {B.~D.}\ \bibnamefont
  {Schultz}}, \ and\ \bibinfo {author} {\bibfnamefont {C.~J.}\ \bibnamefont
  {Palmstr\o{}m}},\ }\href@noop {} {\bibfield  {journal} {\bibinfo  {journal}
  {Surf. Rev. Lett.}\ }\textbf {\bibinfo {volume} {11}},\ \bibinfo {pages}
  {531} (\bibinfo {year} {2004})}\BibitemShut {NoStop}%
\bibitem [{\citenamefont {Leuenberger}\ \emph {et~al.}(2005)\citenamefont
  {Leuenberger}, \citenamefont {Parge}, \citenamefont {Felsch}, \citenamefont
  {Fauth},\ and\ \citenamefont {Hessler}}]{leuenberger2005gdn}%
  \BibitemOpen
  \bibfield  {author} {\bibinfo {author} {\bibfnamefont {F.}~\bibnamefont
  {Leuenberger}}, \bibinfo {author} {\bibfnamefont {A.}~\bibnamefont {Parge}},
  \bibinfo {author} {\bibfnamefont {W.}~\bibnamefont {Felsch}}, \bibinfo
  {author} {\bibfnamefont {K.}~\bibnamefont {Fauth}}, \ and\ \bibinfo {author}
  {\bibfnamefont {M.}~\bibnamefont {Hessler}},\ }\href@noop {} {\bibfield
  {journal} {\bibinfo  {journal} {Phys. Rev. B}\ }\textbf {\bibinfo {volume}
  {72}},\ \bibinfo {pages} {014427} (\bibinfo {year} {2005})}\BibitemShut
  {NoStop}%
\bibitem [{\citenamefont {Bogaerts}\ \emph {et~al.}(1996)\citenamefont
  {Bogaerts}, \citenamefont {Herlach}, \citenamefont {DeKeyser}, \citenamefont
  {Peeters}, \citenamefont {DeRosa}, \citenamefont {Palmstr\o{}m},
  \citenamefont {Brehmer},\ and\ \citenamefont
  {Allen~Jr}}]{bogaerts1996experimental}%
  \BibitemOpen
  \bibfield  {author} {\bibinfo {author} {\bibfnamefont {R.}~\bibnamefont
  {Bogaerts}}, \bibinfo {author} {\bibfnamefont {F.}~\bibnamefont {Herlach}},
  \bibinfo {author} {\bibfnamefont {A.}~\bibnamefont {DeKeyser}}, \bibinfo
  {author} {\bibfnamefont {F.~M.}\ \bibnamefont {Peeters}}, \bibinfo {author}
  {\bibfnamefont {F.}~\bibnamefont {DeRosa}}, \bibinfo {author} {\bibfnamefont
  {C.~J.}\ \bibnamefont {Palmstr\o{}m}}, \bibinfo {author} {\bibfnamefont
  {D.}~\bibnamefont {Brehmer}}, \ and\ \bibinfo {author} {\bibfnamefont
  {S.~J.}\ \bibnamefont {Allen~Jr}},\ }\href@noop {} {\bibfield  {journal}
  {\bibinfo  {journal} {Phys. Rev. B}\ }\textbf {\bibinfo {volume} {53}},\
  \bibinfo {pages} {15951} (\bibinfo {year} {1996})}\BibitemShut {NoStop}%
\bibitem [{\citenamefont {Mostofi}\ \emph {et~al.}(2014)\citenamefont
  {Mostofi}, \citenamefont {Yates}, \citenamefont {Pizzi}, \citenamefont {Lee},
  \citenamefont {Souza}, \citenamefont {Vanderbilt},\ and\ \citenamefont
  {Marzari}}]{MOSTOFI20142309}%
  \BibitemOpen
  \bibfield  {author} {\bibinfo {author} {\bibfnamefont {A.~A.}\ \bibnamefont
  {Mostofi}}, \bibinfo {author} {\bibfnamefont {J.~R.}\ \bibnamefont {Yates}},
  \bibinfo {author} {\bibfnamefont {G.}~\bibnamefont {Pizzi}}, \bibinfo
  {author} {\bibfnamefont {Y.-S.}\ \bibnamefont {Lee}}, \bibinfo {author}
  {\bibfnamefont {I.}~\bibnamefont {Souza}}, \bibinfo {author} {\bibfnamefont
  {D.}~\bibnamefont {Vanderbilt}}, \ and\ \bibinfo {author} {\bibfnamefont
  {N.}~\bibnamefont {Marzari}},\ }\href {\doibase
  https://doi.org/10.1016/j.cpc.2014.05.003} {\bibfield  {journal} {\bibinfo
  {journal} {Comput. Phys. Commun}\ }\textbf {\bibinfo {volume} {185}},\
  \bibinfo {pages} {2309 } (\bibinfo {year} {2014})}\BibitemShut {NoStop}%
\bibitem [{\citenamefont {Julian}(2012)}]{julian2012numerical}%
  \BibitemOpen
  \bibfield  {author} {\bibinfo {author} {\bibfnamefont {S.~R.}\ \bibnamefont
  {Julian}},\ }\href@noop {} {\bibfield  {journal} {\bibinfo  {journal}
  {Comput. Phys. Commun}\ }\textbf {\bibinfo {volume} {183}},\ \bibinfo {pages}
  {324} (\bibinfo {year} {2012})}\BibitemShut {NoStop}%
\bibitem [{\citenamefont {Bogaerts}\ \emph {et~al.}(1993)\citenamefont
  {Bogaerts}, \citenamefont {Van~Esch}, \citenamefont {Van~Bockstal},
  \citenamefont {Herlach}, \citenamefont {Peeters}, \citenamefont {DeRosa},
  \citenamefont {Palmstr\o{}m},\ and\ \citenamefont
  {Allen~Jr}}]{bogaerts1993experimental}%
  \BibitemOpen
  \bibfield  {author} {\bibinfo {author} {\bibfnamefont {R.}~\bibnamefont
  {Bogaerts}}, \bibinfo {author} {\bibfnamefont {A.}~\bibnamefont {Van~Esch}},
  \bibinfo {author} {\bibfnamefont {L.}~\bibnamefont {Van~Bockstal}}, \bibinfo
  {author} {\bibfnamefont {F.}~\bibnamefont {Herlach}}, \bibinfo {author}
  {\bibfnamefont {F.}~\bibnamefont {Peeters}}, \bibinfo {author} {\bibfnamefont
  {F.}~\bibnamefont {DeRosa}}, \bibinfo {author} {\bibfnamefont {C.~J.}\
  \bibnamefont {Palmstr\o{}m}}, \ and\ \bibinfo {author} {\bibfnamefont
  {S.~J.}\ \bibnamefont {Allen~Jr}},\ }\href@noop {} {\bibfield  {journal}
  {\bibinfo  {journal} {Physica B}\ }\textbf {\bibinfo {volume} {184}},\
  \bibinfo {pages} {232} (\bibinfo {year} {1993})}\BibitemShut {NoStop}%
\bibitem [{\citenamefont {Allen~Jr}\ \emph {et~al.}(1991)\citenamefont
  {Allen~Jr}, \citenamefont {DeRosa}, \citenamefont {Palmstr\o{}m},\ and\
  \citenamefont {Zrenner}}]{allen1991band}%
  \BibitemOpen
  \bibfield  {author} {\bibinfo {author} {\bibfnamefont {S.~J.}\ \bibnamefont
  {Allen~Jr}}, \bibinfo {author} {\bibfnamefont {F.}~\bibnamefont {DeRosa}},
  \bibinfo {author} {\bibfnamefont {C.~J.}\ \bibnamefont {Palmstr\o{}m}}, \
  and\ \bibinfo {author} {\bibfnamefont {A.}~\bibnamefont {Zrenner}},\
  }\href@noop {} {\bibfield  {journal} {\bibinfo  {journal} {Phys. Rev. B}\
  }\textbf {\bibinfo {volume} {43}},\ \bibinfo {pages} {9599} (\bibinfo {year}
  {1991})}\BibitemShut {NoStop}%
\bibitem [{\citenamefont {Yang}\ \emph {et~al.}(2017)\citenamefont {Yang},
  \citenamefont {Nummy}, \citenamefont {Li}, \citenamefont {Jaszewski},
  \citenamefont {Abramchuk}, \citenamefont {Dessau},\ and\ \citenamefont
  {Tafti}}]{yang2017extreme}%
  \BibitemOpen
  \bibfield  {author} {\bibinfo {author} {\bibfnamefont {H.-Y.}\ \bibnamefont
  {Yang}}, \bibinfo {author} {\bibfnamefont {T.}~\bibnamefont {Nummy}},
  \bibinfo {author} {\bibfnamefont {H.}~\bibnamefont {Li}}, \bibinfo {author}
  {\bibfnamefont {S.}~\bibnamefont {Jaszewski}}, \bibinfo {author}
  {\bibfnamefont {M.}~\bibnamefont {Abramchuk}}, \bibinfo {author}
  {\bibfnamefont {D.~S.}\ \bibnamefont {Dessau}}, \ and\ \bibinfo {author}
  {\bibfnamefont {F.}~\bibnamefont {Tafti}},\ }\href {\doibase
  10.1103/PhysRevB.96.235128} {\bibfield  {journal} {\bibinfo  {journal} {Phys.
  Rev. B}\ }\textbf {\bibinfo {volume} {96}},\ \bibinfo {pages} {235128}
  (\bibinfo {year} {2017})}\BibitemShut {NoStop}%
\bibitem [{\citenamefont {Tafti}\ \emph {et~al.}(2016)\citenamefont {Tafti},
  \citenamefont {Gibson}, \citenamefont {Kushwaha}, \citenamefont
  {Haldolaarachchige},\ and\ \citenamefont {Cava}}]{tafti2016resistivity}%
  \BibitemOpen
  \bibfield  {author} {\bibinfo {author} {\bibfnamefont {F.~F.}\ \bibnamefont
  {Tafti}}, \bibinfo {author} {\bibfnamefont {Q.~D.}\ \bibnamefont {Gibson}},
  \bibinfo {author} {\bibfnamefont {S.~K.}\ \bibnamefont {Kushwaha}}, \bibinfo
  {author} {\bibfnamefont {N.}~\bibnamefont {Haldolaarachchige}}, \ and\
  \bibinfo {author} {\bibfnamefont {R.~J.}\ \bibnamefont {Cava}},\ }\href@noop
  {} {\bibfield  {journal} {\bibinfo  {journal} {Nat. Phys.}\ }\textbf
  {\bibinfo {volume} {12}},\ \bibinfo {pages} {272} (\bibinfo {year}
  {2016})}\BibitemShut {NoStop}%
\bibitem [{\citenamefont {Kasuya}\ \emph {et~al.}(1996)\citenamefont {Kasuya},
  \citenamefont {Sera}, \citenamefont {Okayama},\ and\ \citenamefont
  {Haga}}]{kasuya1996normal}%
  \BibitemOpen
  \bibfield  {author} {\bibinfo {author} {\bibfnamefont {T.}~\bibnamefont
  {Kasuya}}, \bibinfo {author} {\bibfnamefont {M.}~\bibnamefont {Sera}},
  \bibinfo {author} {\bibfnamefont {Y.}~\bibnamefont {Okayama}}, \ and\
  \bibinfo {author} {\bibfnamefont {Y.}~\bibnamefont {Haga}},\ }\href@noop {}
  {\bibfield  {journal} {\bibinfo  {journal} {J. Phys. Soc. Jpn.}\ }\textbf
  {\bibinfo {volume} {65}},\ \bibinfo {pages} {160} (\bibinfo {year}
  {1996})}\BibitemShut {NoStop}%
\bibitem [{\citenamefont {Sun}\ \emph {et~al.}(2016)\citenamefont {Sun},
  \citenamefont {Wang}, \citenamefont {Guo}, \citenamefont {Liu},\ and\
  \citenamefont {Lei}}]{sun2016large}%
  \BibitemOpen
  \bibfield  {author} {\bibinfo {author} {\bibfnamefont {S.}~\bibnamefont
  {Sun}}, \bibinfo {author} {\bibfnamefont {Q.}~\bibnamefont {Wang}}, \bibinfo
  {author} {\bibfnamefont {P.-J.}\ \bibnamefont {Guo}}, \bibinfo {author}
  {\bibfnamefont {K.}~\bibnamefont {Liu}}, \ and\ \bibinfo {author}
  {\bibfnamefont {H.}~\bibnamefont {Lei}},\ }\href@noop {} {\bibfield
  {journal} {\bibinfo  {journal} {New J. Phys.}\ }\textbf {\bibinfo {volume}
  {18}},\ \bibinfo {pages} {082002} (\bibinfo {year} {2016})}\BibitemShut
  {NoStop}%
\bibitem [{\citenamefont {Li}\ \emph {et~al.}(1996)\citenamefont {Li},
  \citenamefont {Haga}, \citenamefont {Shida}, \citenamefont {Suzuki},\ and\
  \citenamefont {Kwon}}]{li1996electrical}%
  \BibitemOpen
  \bibfield  {author} {\bibinfo {author} {\bibfnamefont {D.~X.}\ \bibnamefont
  {Li}}, \bibinfo {author} {\bibfnamefont {Y.}~\bibnamefont {Haga}}, \bibinfo
  {author} {\bibfnamefont {H.}~\bibnamefont {Shida}}, \bibinfo {author}
  {\bibfnamefont {T.}~\bibnamefont {Suzuki}}, \ and\ \bibinfo {author}
  {\bibfnamefont {Y.~S.}\ \bibnamefont {Kwon}},\ }\href {\doibase
  10.1103/PhysRevB.54.10483} {\bibfield  {journal} {\bibinfo  {journal} {Phys.
  Rev. B}\ }\textbf {\bibinfo {volume} {54}},\ \bibinfo {pages} {10483}
  (\bibinfo {year} {1996})}\BibitemShut {NoStop}%
\bibitem [{\citenamefont {{\.Z}oga{\l}}\ \emph {et~al.}(2014)\citenamefont
  {{\.Z}oga{\l}}, \citenamefont {Wawryk}, \citenamefont {Matusiak},\ and\
  \citenamefont {Henkie}}]{zogal2014electron}%
  \BibitemOpen
  \bibfield  {author} {\bibinfo {author} {\bibfnamefont {O.}~\bibnamefont
  {{\.Z}oga{\l}}}, \bibinfo {author} {\bibfnamefont {R.}~\bibnamefont
  {Wawryk}}, \bibinfo {author} {\bibfnamefont {M.}~\bibnamefont {Matusiak}}, \
  and\ \bibinfo {author} {\bibfnamefont {Z.}~\bibnamefont {Henkie}},\
  }\href@noop {} {\bibfield  {journal} {\bibinfo  {journal} {J. Alloys Compd.}\
  }\textbf {\bibinfo {volume} {587}},\ \bibinfo {pages} {190} (\bibinfo {year}
  {2014})}\BibitemShut {NoStop}%
\bibitem [{\citenamefont {Pavlosiuk}\ \emph {et~al.}(2017)\citenamefont
  {Pavlosiuk}, \citenamefont {Kleinert}, \citenamefont {Swatek}, \citenamefont
  {Kaczorowski},\ and\ \citenamefont {Wi{\'s}niewski}}]{pavlosiuk2017fermi}%
  \BibitemOpen
  \bibfield  {author} {\bibinfo {author} {\bibfnamefont {O.}~\bibnamefont
  {Pavlosiuk}}, \bibinfo {author} {\bibfnamefont {M.}~\bibnamefont {Kleinert}},
  \bibinfo {author} {\bibfnamefont {P.}~\bibnamefont {Swatek}}, \bibinfo
  {author} {\bibfnamefont {D.}~\bibnamefont {Kaczorowski}}, \ and\ \bibinfo
  {author} {\bibfnamefont {P.}~\bibnamefont {Wi{\'s}niewski}},\ }\href@noop {}
  {\bibfield  {journal} {\bibinfo  {journal} {Sci. Rep.}\ }\textbf {\bibinfo
  {volume} {7}},\ \bibinfo {pages} {12822} (\bibinfo {year}
  {2017})}\BibitemShut {NoStop}%
\bibitem [{\citenamefont {Pavlosiuk}\ \emph {et~al.}(2018)\citenamefont
  {Pavlosiuk}, \citenamefont {Swatek}, \citenamefont {Kaczorowski},\ and\
  \citenamefont {Wi\ifmmode~\acute{s}\else
  \'{s}\fi{}niewski}}]{pavlosiuk2018magnetoresistance}%
  \BibitemOpen
  \bibfield  {author} {\bibinfo {author} {\bibfnamefont {O.}~\bibnamefont
  {Pavlosiuk}}, \bibinfo {author} {\bibfnamefont {P.}~\bibnamefont {Swatek}},
  \bibinfo {author} {\bibfnamefont {D.}~\bibnamefont {Kaczorowski}}, \ and\
  \bibinfo {author} {\bibfnamefont {P.}~\bibnamefont {Wi\ifmmode~\acute{s}\else
  \'{s}\fi{}niewski}},\ }\href@noop {} {\bibfield  {journal} {\bibinfo
  {journal} {Phys. Rev. B}\ }\textbf {\bibinfo {volume} {97}},\ \bibinfo
  {pages} {235132} (\bibinfo {year} {2018})}\BibitemShut {NoStop}%
\bibitem [{\citenamefont {Delaney}\ \emph {et~al.}(2010)\citenamefont
  {Delaney}, \citenamefont {Spaldin},\ and\ \citenamefont {Van~de
  Walle}}]{delaney2010}%
  \BibitemOpen
  \bibfield  {author} {\bibinfo {author} {\bibfnamefont {K.~T.}\ \bibnamefont
  {Delaney}}, \bibinfo {author} {\bibfnamefont {N.~A.}\ \bibnamefont
  {Spaldin}}, \ and\ \bibinfo {author} {\bibfnamefont {C.~G.}\ \bibnamefont
  {Van~de Walle}},\ }\href {\doibase 10.1103/PhysRevB.81.165312} {\bibfield
  {journal} {\bibinfo  {journal} {Phys. Rev. B}\ }\textbf {\bibinfo {volume}
  {81}},\ \bibinfo {pages} {165312} (\bibinfo {year} {2010})}\BibitemShut
  {NoStop}%
\bibitem [{\citenamefont {Van~de Walle}\ and\ \citenamefont
  {Neugebauer}(2003)}]{vandewalle2003}%
  \BibitemOpen
  \bibfield  {author} {\bibinfo {author} {\bibfnamefont {C.~G.}\ \bibnamefont
  {Van~de Walle}}\ and\ \bibinfo {author} {\bibfnamefont {J.}~\bibnamefont
  {Neugebauer}},\ }\href@noop {} {\bibfield  {journal} {\bibinfo  {journal}
  {Nature}\ }\textbf {\bibinfo {volume} {423}},\ \bibinfo {pages} {626}
  (\bibinfo {year} {2003})}\BibitemShut {NoStop}%
\end{thebibliography}
%

\end{document}